\RequirePackage[2020-02-02]{latexrelease}
\documentclass[nofootinbib,aps,11pt,twocolumn,letterpaper,preprintnumbers,superscriptaddress,showkeys]{revtex4}
\usepackage{amsmath}
\usepackage{eurosym}
\usepackage{amssymb}
\usepackage{graphicx}
\usepackage{color}
\usepackage{braket}
\usepackage{booktabs}
\usepackage{dcolumn}
\usepackage{multirow}
\usepackage{amsmath}
\usepackage{amssymb}
\usepackage{subfigure}
\usepackage{amsfonts}
\usepackage[left=1.3cm,right=1.3cm,top=1.5cm,bottom=1.5cm]{geometry}
\usepackage[usenames,dvipsnames,svgnames]{xcolor}  %% colored text
\usepackage{hyperref} 
\definecolor{darkgreen}{rgb}{0.1,0.6,0.1}
\definecolor{darkblue}{rgb}{0,0,0.3}
\definecolor{darkred}{rgb}{0.7,0,0}

\definecolor{light gray}{RGB}{220,220,220}
\definecolor{dark purple}{RGB}{108,0,217}
\definecolor{pink}{RGB}{190,20,100}
\definecolor{orang}{RGB}{193,63,0}
\definecolor{green}{RGB}{11,98,17}
\definecolor{darkpink}{RGB}{153,0,76}
\definecolor{bluegreen}{RGB}{0,102,102}
\definecolor{greenlagan}{RGB}{0,102,0}
\definecolor{redgreen}{RGB}{102,102,0}
\definecolor{Redgreen}{RGB}{153,76,0}
\definecolor{vividviolet}{rgb}{0.62, 0.0, 1.0}
\definecolor{amaranth}{rgb}{0.9, 0.17, 0.31}
\definecolor{palatinateblue}{rgb}{0.15, 0.23, 0.89}
\definecolor{brightpink}{rgb}{1.0, 0.0, 0.5}
\definecolor{cornflowerblue}{rgb}{0.39, 0.58, 0.93}
\definecolor{deepcarminepink}{rgb}{0.94, 0.19, 0.22}
\definecolor{radicalred}{rgb}{1.0, 0.21, 0.37}
 
\definecolor{beamer@PRD}{RGB}{46,48,146}
\hypersetup{
     breaklinks=true,
    pdfstartview={FitH},    % fits the width of the page to the window
    colorlinks=true,       % false: boxed links; true: colored links
    linkcolor=blue,          % color of internal links
    citecolor=cyan,        % color of links to bibliography
    filecolor=magenta,      % color of file links
    urlcolor=magenta,           % color of external links
    anchorcolor=cyan,      % Color for anchor text
    linktocpage=true
}

\begin{document}
	{\vskip .1cm}
\date{\today}
\newcommand\be{\begin{equation}}
\newcommand\ee{\end{equation}}
\newcommand\bea{\begin{eqnarray}}
\newcommand\eea{\end{eqnarray}}
\newcommand\bseq{\begin{subequations}} %solo con amsmath
\newcommand\eseq{\end{subequations}}
\newcommand\bcas{\begin{cases}}
\newcommand\ecas{\end{cases}}
\newcommand{\p}{\partial}
\newcommand{\f}{\frac}

\title{Lifetime of scalar cloud formation around a rotating regular black hole}

\author {\textbf{Mohsen Khodadi}}\email{m.khodadi@ipm.ir
}
\affiliation{School of Astronomy, Institute for Research in Fundamental Sciences (IPM),	P. O. Box 19395-5531, Tehran, Iran}
\affiliation{School of Physics, Institute for Research in Fundamental Sciences (IPM),	P. O. Box 19395-5531, Tehran, Iran}
\affiliation{Physics Department, College of Sciences, Shiraz University, Shiraz 71454, Iran}
\affiliation{Biruni Observatory, College of Sciences, Shiraz University, Shiraz 71454, Iran}
\author {\textbf{Reza Pourkhodabakhshi}}
\email{reza.pk.bakhshi@gmail.com
}
\affiliation{Department of Physics, Shahid Beheshti University, G.C., Evin, Tehran 19839, Iran}
\affiliation{Departament de Física Quàntica i Astrofísica, Facultat de Física, Universitat de Barcelona, Martí i Franquès 1, 08028 Barcelona, Spain}
\affiliation{Institut de Ciencies del Cosmos (ICCUB), Universitat de Barcelona, Martí i Franquès 1, 08028 Barcelona, Spain}

 %-------------------------------------------------------------------------------------------------------------------------------------

\begin{abstract}
Does circumventing the curvature singularity of the Kerr black hole affects the timescale of the scalar cloud formation around it? By definition, the scalar cloud, forms a gravitational atom with hydrogen-like bound states, lying on the threshold of a massive scalar field's superradiant instability regime (time-growing quasi-bound states) and beyond (time-decaying quasi-bound states). By taking a novel type of rotating hollow regular black hole proposed by Simpson and Visser which unlike its standard rivals has an asymptotically Minkowski core, we address this question. 
The metric has a minimal extension relative to the standard Kerr, originating from a single regularization parameter $\ell$, with length dimension. We show with the inclusion of the regularization length scale $\ell$ into the Kerr spacetime, without affecting the standard superradiant instability regime, the timescale of scalar cloud formation gets shorter.  Since the scalar cloud after its formation, via energy dissipation, can play the role of a continuum source for gravitational waves, such a reduction in the instability growth time improves the phenomenological detection prospects of new physics because the shorter the time, the more astrophysically important. 


%\vskip -1.0 truecm
 \end{abstract}

\maketitle

\section{Introduction}

\begin{figure}[t]
\begin{tabular}{c}
\includegraphics[width=1\columnwidth]{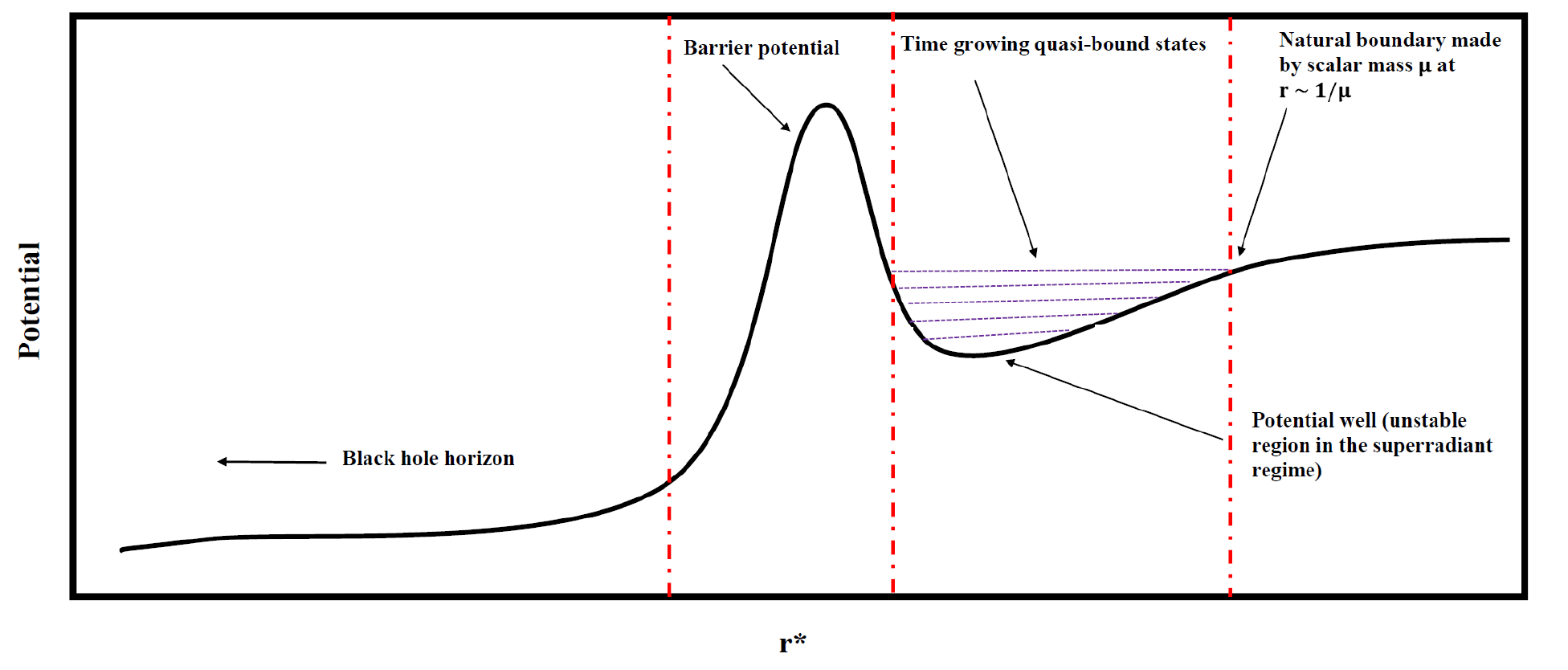}
\end{tabular}
\caption{A schema of the effective gravitational potential with trapping well outside the black hole in which the scalar field with the mass $\mu$, is caught  within it and subsequently grows exponentially with time for creating the instability.}
\label{Gr}
\end{figure}

A reasonable query in connection with the black hole solutions as robust productions of the General Theory of Relativity (GTR) is whether those solutions would be stable long enough to be found in nature. In other words, it is nonsense to look for such entities in nature if they are not stable enough. There are leading studies to support the stability of non-rotating and rotating black hole solutions against gravitational perturbations by massless fields. It has been proven the Schwarzschild metric is  stable against all massless fields \cite{Regge:1957td,Vishveshwara:1970cc,Zerilli:1970wzz}. The Kerr solution as well has been shown to be stable as it gets exposed to massless scalar and gravitational perturbations \cite{Teukolsky:1972my,Press:1973zz}.
Although in the context of studies on black holes' perturbations, there is a phenomenon known as superradiance \cite{Zel:1971,Zel:1972,Starobinsky:1973aij} which complicates the situation somewhat. Concerning this phenomenon, certain perturbations are enhanced by the rotation of a black hole such that the energy radiated away to infinity may exceed the energy of the initial perturbations. Actually, through perturbations vicinity of a Kerr black hole, one can extract its rotational energy \footnote{It is worthy to mention it is not the only way for energy extraction from a black hole, rather there are other theoretical mechanisms such as the Penrose process \cite{Penrose:1969pc} and magnetic reconnection \cite{Comisso:2020ykg,Khodadi:2022dff}. The fact that the superradiance effect, the contrary Hawking radiation \cite{Hawking:1975vcx} is justifiable in the context of classic physics without any origination in quantum mechanics has caused a lot of renewed interest in studying it via employing various classes of black hole solutions admitted by extended theories of gravity (e.g., see \cite{Pani:2011gy}-\cite{Jha:2022ewi}).}. It is shown that superradiance occurs for a perturbed Kerr black hole if the oscillation frequency of the perturbation which is real $\omega_R$, satisfies the inequality $\omega_R<m \Omega_+$. Namely, the oscillation frequency of the perturbations should be smaller than the product of the azimuthal number of the perturbation $m$ and the angular velocity on the event horizon $\Omega_+$.
By referring to seminal paper \cite{Press:1972zz}, one will face the postulation stating that if the superradiance arising from a perturbed black hole becomes reflected toward the event horizon of the black hole repeatedly; then, due to an initial small perturbation, an exponential growth without any bound arises which results in a new phenomenon known as black hole bomb \cite{Cardoso:2004nk}.  The mentioned effect is one of the most considered processes around black holes, whose key characteristic is trapping the radiation between the event horizon and a reflecting surface outside the black hole. The mechanism of reflection toward a black hole is commonly created by the rest mass of the perturbing field naturally, and either artificially by a reflector surface such as a mirror or cavity, see Refs. \cite{Furuhashi:2004jk}-\cite{Vieira:2021nha}. The asymptotically Anti-de-sitter (AdS) spacetime is also a natural bed to make the reflecting boundary for waves since the spatial infinity is at a finite distance causally, resulting in the formation of a time-like boundary \cite{Cardoso:2006wa}-\cite{Rahmani:2020vvv}.
In this way, a spectrum of quasi-bound states (QBS) forms in the potential well outside the black hole so that it disappears at spatial infinity \cite{Grain:2007gn,Hod:2017gvn}. To better visualize the black hole bomb phenomenon, the above-mentioned explanations are summarized in Fig. \ref{Gr} schematically. 

The terminology of QBS comes from the fact that the bound states formed within the black hole's potential well have complex frequencies $\omega_R+i\omega_I$ so that its imaginary part denotes the decay and either growth rate of the perturbation with time. So, we deal with two types of configurations: time-decaying QBSs and time-growing QBSs. The former is the general behavior expected for the matter surrounding a black hole (particularly Schwarzschild) due to the purely ingoing boundary condition at the event horizon. The latter is devoted to the rotating black hole in the superradiant regime $\omega_R<m\Omega_+$, which yields an instability in the spinning black hole in the presence of any field which lets QBS form in the superradiant regime.  In literature, the use of the term superradiant bound state (SBS) for the  time-growing QBSs is common, too \cite{Baryakhtar:2020gao}. Note the superradiant scattering singly does not guarantee the superradiant instability, rather the presence of QBSs in the superradiant regime (time-growing modes) is an essential condition.
Actually, may the superradiant scattering occur but the superradiant instability not, see Refs. \cite{Detweiler:1973zz,Hod:2012wmy,Hod:2013nn} (also \cite{Furuhashi:2004jk}).  Inasmuch as the superradiance effect occurs only for the bosonic fields with integer spin \cite{Brito:2015oca}, typically, people are interested in superradiant instability triggered by the massive scalar fields \footnote{In the meantime some interesting studies can be found on higher spin massive fields, as well \cite{Sampaio:2014swa}-\cite{East:2018glu}. The reason for such much attention to the scalar field is multifaceted. First of all, the scalar fields are ubiquitous in theoretical physics (Higgs boson is the most famous case), and may represent a fundamental degree of freedom. It is well-known, due to the capability of these entities to give rise to long-range gravitational fields, they are suitable candidates for building up extended theories of gravity. Scalar fields are of undeniable importance in a fundamental theory such as string theory. Scalar field dark matter models with the assumption of a very tiny mass corresponding to the scalar field are among the favorite candidates to solve one of the biggest challenges in explaining the mechanism governing the Universe.	It is also essential to mention one of the cornerstones for the development of modern cosmology is the so-called scalar field inflation which is a successful theory in predicting descriptions compatible with observations from the early Universe (without addressing the initial singularity issue).}.  
As an interesting point, the recent studies indicate the shadow recorded by the Event horizon Telescope \cite{EventHorizonTelescope:2019dse} can be served to shed light on the superradiant instability phenomenon, see Refs. \cite{Creci:2020mfg,Roy:2021uye} (see also \cite{Roy:2019esk, Cunha:2019ikd}).

Altogether, perturbating the Kerr black hole by the massive scalar field, in the case of satisfying the superradiance regime, results in transferring the energy and the angular momentum of the black hole to the scalar field until reaching to the saturation point $\omega_R\sim m\Omega_+ \sim \mu$. Because of the energy extraction, it is expected to form an equilibrium configuration of a complex massive scalar field in the Kerr background; so-called scalar clouds. More technically, scalar clouds are bound states that there exist at the threshold of two regimes: SBSs (time-growing QBSs) and beyond (time-decaying QBSs) \cite{Sampaio:2014swa}. Indeed, superradiant instability in the background of a black hole results effectively in the formation of a gravitational atom, featuring hydrogen-like states between the black hole and the scalar bosonic cloud.
A significant point in connection with scalar cloud formation (SCF) is the characteristic timescale to reach the saturation point, $\tau_{nlm}\equiv 1/\omega_I$. In other words, $\tau_{nlm}$ is the time required for a scalar cloud to reach to its maximum mass, i.e., the final stage of the formation.
The rotating scalar bosonic cloud is of phenomenological importance in the sense that after its formation, on very long timescales, it dissipates its energy through the emission of monochromatic gravitational waves within a very narrow range of frequencies around $\omega_R/\pi\approx \mu/\pi$ \cite{Yuan:2021ebu}. This lets us probe the imprint of the scalar bosonic field via gravitational waves since the scalar cloud's configuration acts similar to a continuous gravitational wave source (e.g., see \cite{Arvanitaki:2014wva}-\cite{Baumann:2022pkl}).

By taking the astrophysical considerations for the no-hair theorem, the unique solution for a steady black hole in four-dimensional vacuum of GTR is the Kerr black hole \footnote{This is also known as the Kerr hypothesis, meaning all astrophysical black holes are well-described by Kerr geometry.  Although in the light of the two images recorded by the Event Horizon Telescope, we expect the galaxies Messier 87 \cite{EventHorizonTelescope:2019pgp} and Milk Way  \cite{EventHorizonTelescope:2022xnr} to host the supermassive Kerr black holes M87* and Sgr A* respectively; concerning the universality of this hypothesis, there are some theoretical conflicts, see \cite{Herdeiro:2022yle} for more details.}.  However, the existence of singularities in the form of a ring in the center of the Kerr spacetime has no meaning but the failure of the GTR, since they denote a region at which the causal connection of geodesics disrupts suddenly. Namely, the spacetime singularities, in essence, result in losing the predictive power of theory in very small scales. 
So, the principal motivation behind constructing non-singular alternatives to Kerr black holes (regardless of their type) comes from the widespread belief that singularities merely are nonphysical objects imposed by the classical theories of gravity \cite{Hawking} and should not exist in nature. Theoretically, we have to see Kerr black hole merely as an effective representation of a more comprehensive metric. Although it is believed taking quantum gravity considerations into account can fix the issue, though, such a definite theory is still out of reach. 
Nevertheless, the efforts did not stop and instead have been conducted remarkably in the direction of constructing the regular models (also named non-singular black holes). Namely, models representing black holes with a non-singular core effectively. Historically, the idea of regular black holes dates back to the seminal papers \cite{Sakharov:1966aja,Gliner}. The key idea stated if the interior of a black hole fills with some matter with the equation of state $\rho=-p$, i.e., corresponding to a de Sitter (dS) core, then, the singularity in the center is circumvented. In other words, the negative pressure arising from the dS core prevents geodesics from reaching the singularity point. Most of the regular black holes were originated from the original Bardeen model \cite{Bardeen} until other models, e.g., \cite{Bogojevic:1998ma}-\cite{Toshmatov:2014nya}, were built by the inspiration of this idea. The price paid to bypass the singularity in these models with dS core (usually named standard regular black holes) is the violation of the weak/strong energy conditions. Studies indicate the physical source of the standard regular black holes should be found in the nonlinear electromagnetic \cite{Fan:2016hvf}. Although the development of the standard electromagnetic to a nonlinear one is theoretically well-motivated, so far, the observations have not given us a signal for the existence of such a magnetic charge in nature.

In this paper, we desire to investigate the role of the avoidance from the central singularity of Kerr black hole on the lifetime of SBSs triggered by a massive scalar field.  Even though the singularity is hidden within the event horizon, we are interested to see whether removing the singularity of the Kerr black hole affects the growth time of QBSs (the timescale of SCF around the black hole). Given the phenomenological prospect of SCF, this study is astrophysically well-motivated, since we see by imposing a natural and inevitable assumption (avoiding curvature singularity) into the standard Kerr metric, it results in a shorter timescale of SCF. To do so, we adopt a composite system including a massive scalar field propagating on a novel type of rotating hollow regularized black hole which unlike standard regular models has an asymptotically Minkowski core. This metric is known as Simpson-Visser (SV), and its static spherically symmetric solution, first, has been released in \cite{Simpson:2018tsi}; and its rotating version can be found in \cite{Simpson:2021dyo,Simpson:2021zfl} as well. 
The notable point distinguishing the SV metric from other standard regular models is the fact that it was built based on a set of theoretical and observational motivated criteria. In this regard, of the salient features which can be mentioned is the SV metric models a spinning black hole which is regular everywhere; it recovers Kerr at large distances and has integrable geodesics. Besides, within the range of theoretical and observational verification, fulfills all of the energy conditions admitted by GTR. 
Due to the existence of an asymptotically Minkowski core in the SV metric, it indeed is a hollow black hole with an interior region's physics being much simpler in comparison with the standard ones \cite{Simpson:2019mud}. This metric is rich geometrically, in the sense that it potentially is prone to address compact objects other than a black hole, too \cite{Simpson:2019cer}. ‌Besides, in the light of the separability of the Klein–Gordon (KG) equation as well as Maxwell’s equations on the SV spacetime, it can be considered as a well-behaving background to analyze the quasi-normal modes associated with the spin-zero and spin-one perturbations \cite{Simpson:2021biv,Franzin:2022iai}. These tempting features have caused the hollow rotating regular black hole at hand, to become a more ideal black hole astrophysically relative to standard regular models. This has led to a lot of attention being paid to its various physical aspects (see e.g., Refs. \cite{ Bargueno:2020ais}-\cite{Patel:2022jbk}).  

In this regard, by taking the SV rotating regular black hole as a framework, we organized this paper as follows. After a short review of the underlying metric in Sec. \ref{SV}, we go to Sec. \ref{SIR} for the derivation of the superradiant instability regime triggered by massive scalar perturbation around SV rotating regular black hole. In Sec. \ref{QBS}, by analyzing the SBSs, we investigate the effect of regularization of the central singularity in Kerr black hole (as proposed by SV) on the lifetime of SCF. A summary of the results can be found in Sec. \ref{cons}. Throughout this paper, we use the signature convention $(-,+,+,+)$ and for simplicity, we work with the units $c=G_N=\hbar=1$.
%%%%%%%%%%%%%%%%%%%%%%%%%%%%%%%%%%%%%%%%%%%%%%%%%%%%%%%%%%%%%%%%%
\section{rotating regular SV black hole}
\label{SV}
Recently, by aiming to cancel out singularity, SV have proposed a static and spherically symmetric metric with a non-singular modification to Schwarzschild \cite{Simpson:2019mud}
\begin{eqnarray}\label{AM}
d s^2& =&-\left(1-\frac{2M\,e^{-\ell/r}}{r}\right)\,d t^2 + \frac{d r^2}{1-\frac{2M\,e^{-\ell/r}}{r}} + \\ \nonumber
&&r^2\,d\theta^2+
r^2sin^2\theta d\phi^2 \ ,
\end{eqnarray}
where $M$ and $\ell$ are respectively, ADM mass and a positive suppression parameter with the dimension of length to control the central singularity at $r=0$. The parameter $\ell$, is also named the regularization length scale.
By serving the Newman-Janis algorithm, the rotating version of the metric (\ref{AM}), in the standard Boyer-Lindquist coordinates $(t,r,\theta,\phi)$, is written as
\footnote{A notable point is metric (\ref{eos}) is nothing but the same geometry introduced by Ghosh \cite{Ghosh:2014pba}, with a difference that  SV have presented remarkably more detailed physical discussions and arguments to explain why it is of our interest.} \cite{Simpson:2021dyo,Simpson:2021zfl}
\begin{eqnarray}\label{eos}
d s^2 &=&-\big(\frac{\Delta_{SV}-a^2\sin^2\theta}{\Sigma}\big) d t^2
+\frac{\Sigma}{\Delta_{SV}}\,d r^2 + \\ \nonumber
&&\Sigma\,d\theta^2 + \big(\frac{\sin^2\theta((r^2+a^2)^2-\Delta_{SV}a^2\sin^2\theta)}{\Sigma}\big)d\phi^2+\\ \nonumber
&&\big(\frac{2 a\sin^2\theta((r^2+a^2)-\Delta_{SV})}{\Sigma}\big)dtd\phi
\ ,
\end{eqnarray}
where
\begin{equation}
\Sigma = r^2+a^2\cos^2\theta \ , \qquad \Delta_{SV}=r^2+a^2-2Mr\,e^{-\ell/r} \ .
\end{equation}
By relaxing the single suppression parameter $\ell$ in the metric above, the standard Kerr spacetime is restored; i.e., in the above metric, $\Delta_{Kerr}=r^2+a^2-2Mr$ instead of $\Delta_{SV}$. As another consistency check with Kerr metric, it can be seen as $r\rightarrow+\infty$, asymptotic flatness is preserved. Despite the fact that similar to the standard Kerr, the domains for the temporal and angular coordinates are unaffected, the discontinuity at $r=0$, is limited just to $r\geq0$. This makes the SV metric rich in the sense that the closed time-like curves no longer appear in the surrounding maximally extended Kerr. The metric enjoys a special feature: the ring singularity is replaced by a region of spacetime that is asymptotically Minkowski. It is special because in the standard regular black holes, the ring singularity is replaced by a region of spacetime which is asymptotically dS. 

To consider the regularity problem of the metric at hand, it is just sufficient for us to investigate the behavior of curvature invariants $R_{\alpha\beta} R^{\alpha\beta}$ and $R_{\alpha\beta \mu\nu} R^{\alpha\beta \mu\nu}$, as $r\longrightarrow0$. By taking the equatorial plane ($\theta=\pi/2$), these two invariant read as 
\begin{align}
&R_{\alpha\beta}R^{\alpha\beta} = \frac{32M^2\,e^{-4 \Xi}}{r^6}\big(\Xi^4-2\Xi^3+2\Xi^2\big) \ ,\\
&R_{\alpha\beta \mu\nu} R^{\alpha\beta \mu\nu} = \frac{48M^2\,e^{-4 \Xi}}{r^6}\Big( \frac{4}{3}\Xi^4-\frac{16}{3}\Xi^3+8\Xi^2-4\Xi+1\Big)
\end{align}
where $\Xi=\ell/2r$. As it is clear, both curvature invariants are regular everywhere, including at $r=0$. In this manner, the SV spacetime models a tractable rotating regular black hole with an asymptotically Minkowski core whose geometry is non-singular globally.

Concerning each one of the horizons' exact locations (inner $r_-$ and outer $r_+$) corresponding to the underlying spacetime, we have to solve the roots of $g_{rr}=0$ or $\Delta_{SV}(r)$ analytically.
Despite the fact that $\Delta_{SV}(r)=0$ is not analytically solvable, by adopting a realistic approximation, it turns possible. By introducing a dimensionless regularization parameter $\epsilon=\ell/M$, one of the possible approximated solutions is to write Taylor series expansion about small $\epsilon$. 
%It is a relevant approximation since it is expected for the regularization length scale  $\ell$ to be around Planck length $l_p$.
As an outcome, up to the second-order $\mathcal{O}(\epsilon^2)$, we find
\begin{equation}\label{p}
\frac{r_{\pm}}{M} = 1 \pm\sqrt{1-\chi^2-2\epsilon-\mathcal{O}(\epsilon^2)} \ ,
\end{equation} where $\chi=a/M$ is the dimensionless spin parameter.
To obtain another relevant approximation instead of expanding around $\epsilon=0$, one can look for the approximated horizons' location by expanding about the Kerr horizon located at $\frac{r_{\pm,Kerr}}{M}=1\pm\sqrt{1-\chi^2}$. By Keeping the terms up to the second-order $\mathcal{O}(\epsilon^2)$, we have the following expressions
\begin{align}\label{k}
\frac{r_{\pm}}{M} &= 1 \pm\sqrt{1-\chi^2}~ \mp \\ \nonumber
& \frac{2\sqrt{1-\chi^2}\pm(2-\chi^2)}{(\sqrt{1-\chi^2}\pm 1)\left(\sqrt{1-\chi^2}\pm(1-\chi^2)\right)}\epsilon + \mathcal{O}\left(\epsilon^2\right)	\ ,
\end{align} for the location of outer $r_+$ and inner $r_-$ horizons, respectively.
In this regard, by taking the above approximations into the account of $g_{tt}=0$, the ergosurface is characterized by the following expressions respectively,
\begin{equation} \label{erg1}
\frac{r_{\text{erg}}}{M} = 1 + \sqrt{1-\chi^2\cos^2\theta-2\epsilon-\mathcal{O}(\epsilon^2)} \ ,
\end{equation}
and
\begin{equation}\label{erg2}
\frac{r_{\text{erg}}}{M} = \frac{r_{\text{erg,Kerr}}}{M} - \frac{\frac{2 r_{\text{erg,Kerr}}}{M}-\chi^2\cos^2\theta}{\frac{r_{\text{erg,Kerr}}}{M}\left(\frac{r_{\text{erg,Kerr}}}{M}-\chi^2\cos^2\theta\right)}\,\epsilon + \mathcal{O}(\epsilon^2) \ ,
\end{equation} where $\frac{r_{\text{erg,Kerr}}}{M}=1+\sqrt{1-\chi^2\cos^2\theta}$. 
Considering the fact that the given approximate procedures used for deriving the solutions (\ref{p})-(\ref{erg2}), i.e., the expansions around small values of $\epsilon$ and/or around the standard Kerr, are equivalent in essence; thus, one has to carefully set the values small enough for $\epsilon$. Otherwise, the solutions (\ref{p}) with (\ref{k}) and (\ref{erg1}) with (\ref{erg2}),  may no longer match, and subsequently address distinct locations for $r_{\pm}$ and $r_{\text{erg}}$.

Given the important role of the suppression parameter $\ell$ throughout the analysis, it is interesting to note from  Ref. \cite{Simpson:2021dyo} remember that one has the freedom to determine its scale. More precisely, Simpson and Visser based on the current belief of relativists, categorized the domain of validation for the GTR into the following three domains:
\begin{itemize}
\item Everywhere except for small scales where a phenomenological theory of quantum gravity is essential.
\item Everywhere outside any Cauchy horizon(s).
\item Outside any horizon but has a full stop.
\end{itemize}	
While the first restricts the scale of $\ell$ to small ones, namely the Planck length $l_p$, the remaining two allow us to go beyond the Planck scale.

A profoundly desirable property of the SV spacetime is its connection with the classical energy conditions admitted in GTR. In this respect, the stress-energy components of SV metric takes the following forms
\begin{eqnarray}
	\rho = -p_{r} = \frac{\ell M\,e^{-2\Xi}}{4\pi\Sigma^2} \ , ~~~
	p_{t} = \frac{\ell M\,e^{-2\Xi}}{4\pi\Sigma^2}\left(1-\Xi\right) \ .
\end{eqnarray}
At first sight, globally it satisfies $\rho>0$, if $\ell>0$. Considering $\rho+p_{r}=0$ being valid globally, the radial null energy condition would be satisfied then. Concerning the transverse null energy condition $\rho + p_{t}>0$, one can easily show in the equatorial plane, the violated region is pushed into $r<\frac{\ell}{4}$. Moreover, for the strong energy condition $\rho+p_{r}+2p_{t}>0$, in the equatorial plane, the violated region is absorbed into $r<\frac{\ell}{2}$. If we take the suppression parameter $\ell$ to be of the order of magnitude of a tiny scale \footnote{ As an example, by adopting the above-mentioned first category if we set $\ell \sim l_p$, the regions whose energy condition violates physics will theoretically be pushed into a forbidden region since Planck length $l_p$ is the smallest scale in nature. So, the existence of such forbidden regions $r<l_p/4$ and $r<l_p/2$ for energy condition's violation are not logically and practically a threat to spacetime.}, then the above results indicate all of the violations of energy conditions in physics can be pushed into an extremely small region in the deep core of spacetime which is beyond the reach of the observer.

To find out more details about the analysis of the metric at hand, moreover Ref. \cite{Simpson:2021dyo} we also refer the reader to Refs. \cite{Simpson:2021zfl,Simpson:2019cer,Simpson:2019mud}. 
%%%%%%%%%%%%%%%%%%%%%%%%%%%%%%%%%%%%%%%%%%%%%%%%%%%%%%%%%%%%%%%%%%%%%%%%%%
\section{Analyzing the superradiant instability regime of massive scalar fields using Trapping well} \label{SIR}

In this section, by taking the black hole bomb mechanism \cite{Press:1972zz} into the account of the rotating regular black hole metric \eqref{eos}, we desire to investigate the role of the circumvention of singularity on superradiant instability regime. The existence of a potential well outside the black hole in addition to the existence of the ergo-region is required \cite{Hod:2012zza}. In other words, when the massive modes corresponding to a composited system of the rotating regular background (\ref{eos}) and massive scalar perturbations $\psi$, get stuck in the effective potential well outside the black hole, it may result in creating instability in the system; as schematically shown in Fig. \ref{Gr}. As noted already, these massive modes play the role of a reflecting surface, like a mirror, which causes the wave to be enclosed between itself and the black hole. As a result of the resonance arising from forward and backward movements of the enclosed wave around the black hole, it induces a superradiant instability known as the black hole bomb.

Beginning from the KG equation for the scalar field $\psi$ with the mass $\mu$, we have
\begin{equation}\label{eq:KGEq}
\left(\nabla_\alpha \nabla^\alpha +\mu^2\right)\, \Psi(t,r,\theta,\phi)=0~.
\end{equation}
By means of introducing the following ansatz in the Boyer-Lindquist coordinates 
\begin{eqnarray}\label{eq:an}
\Psi(t,r,\theta,\phi)
&=& R_{lm}(r)\; Y_{lm}(\theta)\; e^{-i\omega t}\; e^{+i m\phi }~,\nonumber \\
&&l\geq0,~~~~l\leq m\leq l,
\end{eqnarray} the function of the scalar field can be decomposed in terms of 
the radial function $R_{lm}(r)$, and the spherical wave function $ Y_{lm}(\theta)$. The subscripts $l,~m$, respectively are the angular quantum number, the azimuthal wave number, and $w$ represents the frequency of the scattering scalar field which is positive ($w>0$). By carrying the components of the metric (\ref{eos}) into the differential equation \eqref{eq:KGEq} along with inserting the ansatz \eqref{eq:an}, the radial KG equation, after separation, reads as
\begin{equation}\label{eq:kg}
\Delta_{SV}{{d}\over{dr}}\Big(\Delta_{SV}{{dR_{ lm}}\over{dr}}\Big)+\mathcal{U} R_{ lm}=0~,
\end{equation}
where 
\begin{equation}\label{eq:M}
\mathcal{U}_{SV}\equiv \bigg((r^2+a^2)\omega-ma\bigg)^2+\Delta_{SV}\bigg(2ma\omega-a^2\omega^2-l(l+1)- \mu^2 r^2\bigg)~.
\end{equation}
To address the black hole bomb phenomena via the radial KG equation \eqref{eq:kg}, we should set the following solution
\begin{eqnarray}\label{eq:so}
R_{lm}\sim\left\{
\begin{array}{ll}
e^{-i (\omega-m\Omega_+)r_*}\ \ \text{ as }\ r\rightarrow r_{+}\ \
(r_*\rightarrow -\infty)  \\\\
\frac{e^{-\sqrt{\mu^2-\omega^2}r_*}}{r}\ \ \text{ as }\
r\rightarrow\infty\ \ \ \ \ \ (r_*\rightarrow +\infty)
\end{array}
\right.
\end{eqnarray} where $\Omega_+=\frac{a}{2Mr_+}=\frac{a}{r_+^2+a^2}$, is angular velocity of a rotating black hole. Note in the above solution $r_*$ is tortoise radial coordinate, coming from the definition $\frac{dr_*}{dr}=\frac{r^2+a^2}{\Delta_{SV}}$. Solution \eqref{eq:so}  asserts the fact that the scalar wave on the horizon is purely ingoing; while, in case of $\omega^2<\mu^2$, it is a bounded solution at spatial infinity, i.e., decaying exponentially.

Now, by suggesting a new radial function as $\psi_{lm}\equiv \sqrt{\Delta_{SV}}R_{lm}$, into the radial equation (\ref{eq:kg}), we come to the following equation known as Regge-Wheeler 
\begin{equation}\label{eq:RW}
\bigg({{d^2}\over{dr^2}}+\omega^2-V\bigg)\psi_{lm}=0\  ,
\end{equation}
with
\begin{equation}\label{eq:RW2}
V=\omega^2-{{f+\mathcal{U}_{SV}}\over{\Delta_{SV}^2}}\,,
\end{equation} where 
\begin{eqnarray}\label{}
f&=&a^2 \ell^2 M r^{-3}e^{-\ell/r}-a^2+ \\ \nonumber
&&M r^{-2} e^{-\frac{2 \ell}{r}}  \bigg((\ell^2-2 \ell r)(r e^{\ell/r}-M)
+M r^2\bigg)~. 
\end{eqnarray} 
It is not difficult to show by putting $\ell=0$ in the above expression, the equation (\ref{eq:RW2}) recovers its standard form
\begin{equation}
V=\omega^2-{{M^2-a^2+\mathcal{U}_{Kerr}}\over{\Delta_{Kerr}^2}}\,,
\end{equation}

By choosing the change of variable $r\longrightarrow x^{-1}$, the effective potential (\ref{eq:RW2}), re-writes as
\begin{widetext}
\begin{eqnarray}
V&=&\frac{-1}{\big(\left(a^2 x^2+1\right) e^{\ell x}-2 M x\big)^2}\Bigg(
M x \ell^{\ell x} \bigg(a^2 \ell^2 x^6+\omega^2 \left(6 a^2 x^2+4\right)-4 m a\omega x^2+\ell^2 x^4-\bigg. \\ \nonumber
&&\bigg. 
2 \ell x^3+2 l(l+1) x^2+2 \mu ^2\bigg)-
e^{2 \ell x}
\bigg(l(l+1) \left(a^2 x^4+x^2\right)+a^2 x^2 \left(\mu^2+\omega^2+x^2(1-m^2)\right)+\bigg. \\ \nonumber
&&\bigg.
a^4 \omega^2 x^4+\mu
^2\bigg)-
 \left(M^2 x^4 \left(\ell^2 x^2-2 \ell x-1\right)+4 M^2w^2 x^2\right)\Bigg)
\end{eqnarray}
\end{widetext}
By expanding around $x=0$, and keeping the terms up to $\mathcal{O}(x^3)$, we can show the asymptotic form of the effective potential $V$ (i.e. $V(r\longrightarrow\infty)$), takes the following form 
%\begin{widetext}
\begin{eqnarray}\label{ex}
&&\mu ^2+\left(2 \mu ^2 M-4 M \omega^2\right)x+ \Bigg(a^2 (\omega^2-\mu ^2)+ \Bigg. \\ \nonumber
&&\Bigg.
(4 M \omega^2-2 \mu ^2M)\ell-12 M^2 \omega^2+
l(l+1)+4 \mu ^2 M^2\Bigg)x^2+\\ \nonumber
&&\mathcal{O}(x^3)
\end{eqnarray}
%\end{widetext}
One immediately realizes if we keep the terms up to $\mathcal{O}(x^2)$ in (\ref{ex}), we are dealing with quite the same thing we expect for the standard Kerr. Given the fact that in black hole bomb phenomenon, the existence of a trapping potential well is essential to have instability; thus, we demand the asymptotic derivative of the effective potential to be positive, i.e., $dV/dr\to 0^+$ as $r\to\infty$ \cite{Hod:2012zza}. By bringing $dV/dr$ out for the expansion (\ref{ex}), we have
\begin{equation}\label{eq}
\frac{(2M\omega^2-M\mu^2)(r-2\ell)}{r^3}-\dfrac{a^2(\omega^2-\mu^2)-12M^2\omega^2+l(l+1)}{r^3}\,.
\end{equation}
Considering the fact that we are interested in the asymptotic derivative of the effective potential, such that $r\gg \ell$ ($r-2\ell \approx r$); it causes the first expression to be prevailed in (\ref{eq}). As a consequence, if 
\begin{equation}\label{eq:reg}
\frac{\mu^2}{2}<\omega^2<\mu^2\,,
\end{equation} the condition $dV/dr>0$ satisfies.
Mixing the above instability regime with the superradiant condition $\omega<m\Omega_+$,  one yields the superradiant instability regime  
\begin{equation}\label{eq:st}
\mu<\sqrt{2}m\Omega_+\,,
\end{equation} which has nothing new but the same instability regime related to a composed system of the massive scalar field and the standard Kerr BH.
Likewise, we can see by keeping the terms up to $\mathcal{O}(x^3)$, including the suppression parameter $\ell$, the standard superradiant instability regime, holds yet. So, fixing the core singularity by an asymptotically Minkowski core characterized by the single suppression parameter $\ell$, has no effect on the standard superradiant instability regime.

\section{Effect of regularization parameter $\ell$ on the lifetime of SBS}
\label{QBS}
The aim in this section is to find the role of the regularization parameter $\ell$ on the lifetime of SBS (or time-growing QBSs). For the first time, the lifetime of SBS around the standard Kerr black hole using the semi-analytical method introduced in the seminal paper \cite{Starobinsky:1973aij}, was derived in Ref. \cite{Detweiler:1980uk}. 
The meaning of the semi-analytical method is analytical asymptotic matching, here. In this method, instead of extracting a direct solution from KG, one indeed obtains a matched solution of the given solution close to the event horizon ($r-r_+\ll \omega^{-1}$) and the one far away from the event horizon ($r-r_+\gg M$). As a result, this solution just is valid in the overlapping range $M\ll r-r_+\ll\omega^{-1}$.  So, the analytical asymptotic matching method performs by hiring two approximations: $M\mu\ll1$ and $M\omega\ll1$, i.e., the product of the mass of the field $ \mu$ and the perturbation frequency $\omega$ with the black hole mass $M$ separately, are much less than the unity. 
The former requires the scalar boson’s Compton wavelength to be bigger than the black hole's gravitational radius; while the latter comes from a low energy regime for perturbation. Multiplication $M\mu$, in  essence, is a criterion for the measurement of the gravitational coupling between the perturbations and the black hole, so that if $M\mu \ll 1$, it gives us the exciting phenomenological message stating that in the light of superradiant instability, the black holes with masses $M/M_\odot$ in the range $\mathcal{O}(1)-\mathcal{O}(10^{10})$
may be utilized to probe the ultra-light scalar bosons with masses $\mu/eV$ in the range $\mathcal{O}(10^{-22})-\mathcal{O}(10^{-10})$.
\begin{figure*}[ht!]
\begin{tabular}{c}
%\centering
\includegraphics[width=0.65\columnwidth]{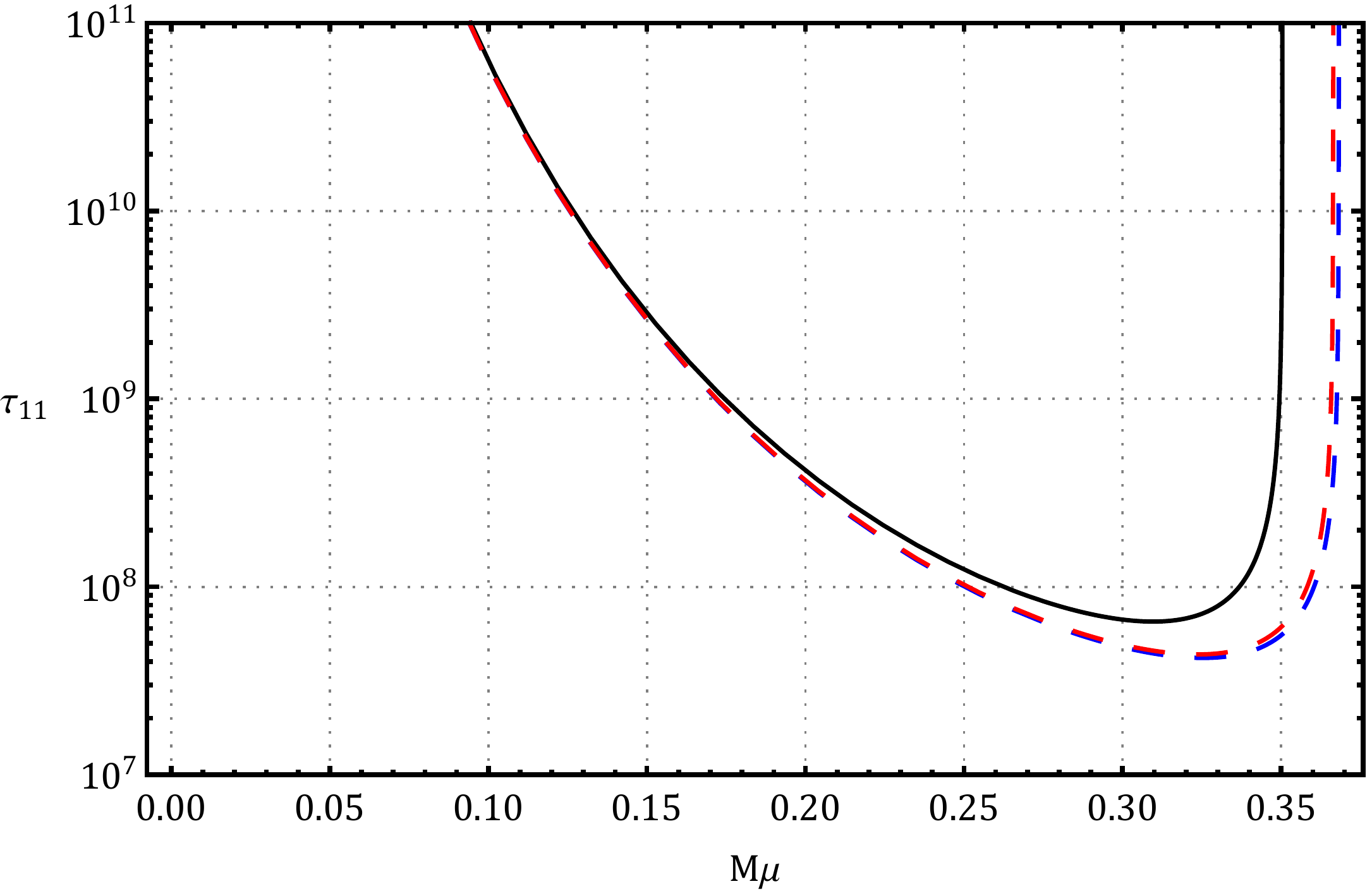}~~
\includegraphics[width=0.65\columnwidth]{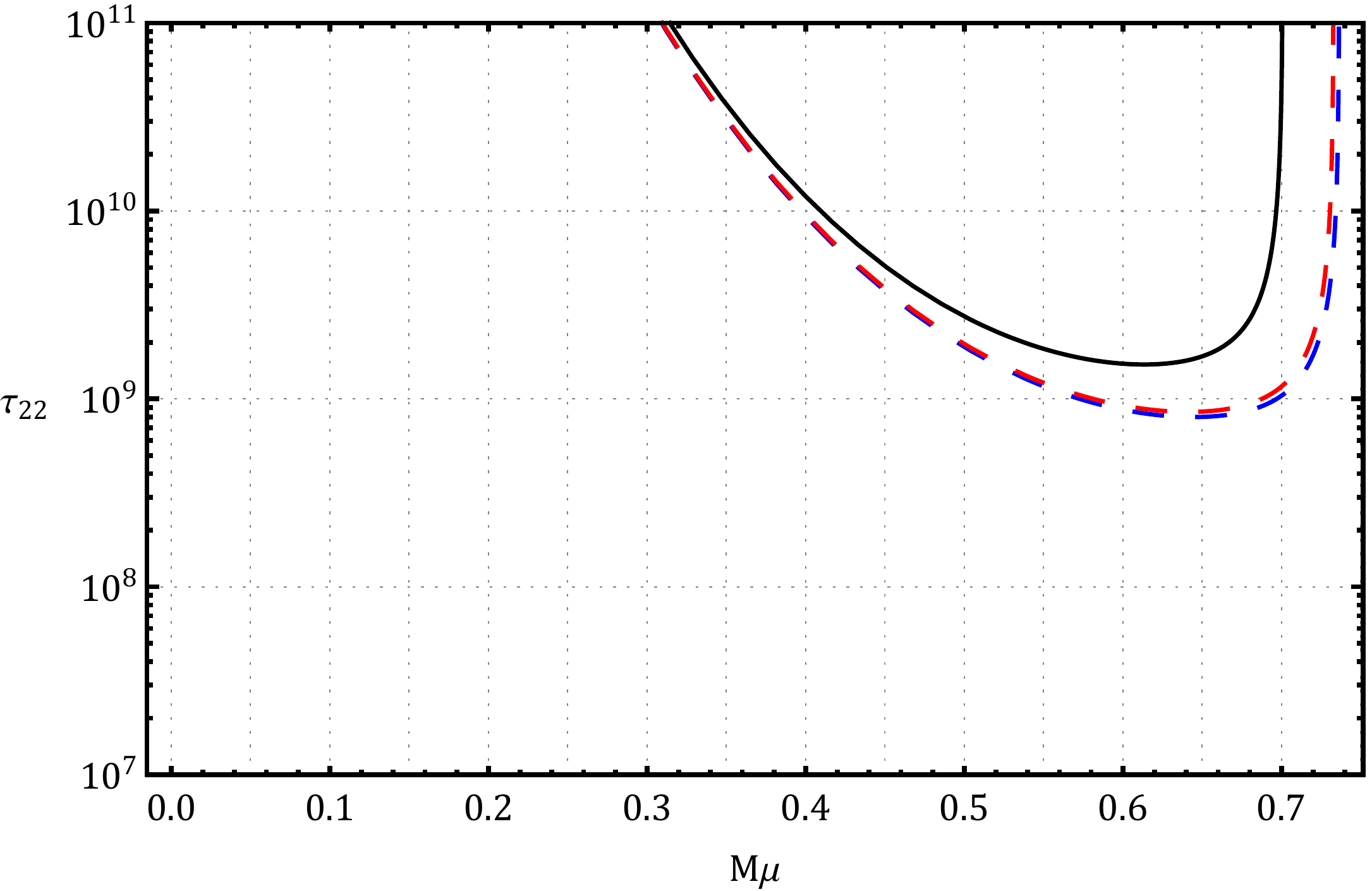}~~
\includegraphics[width=0.65\columnwidth]{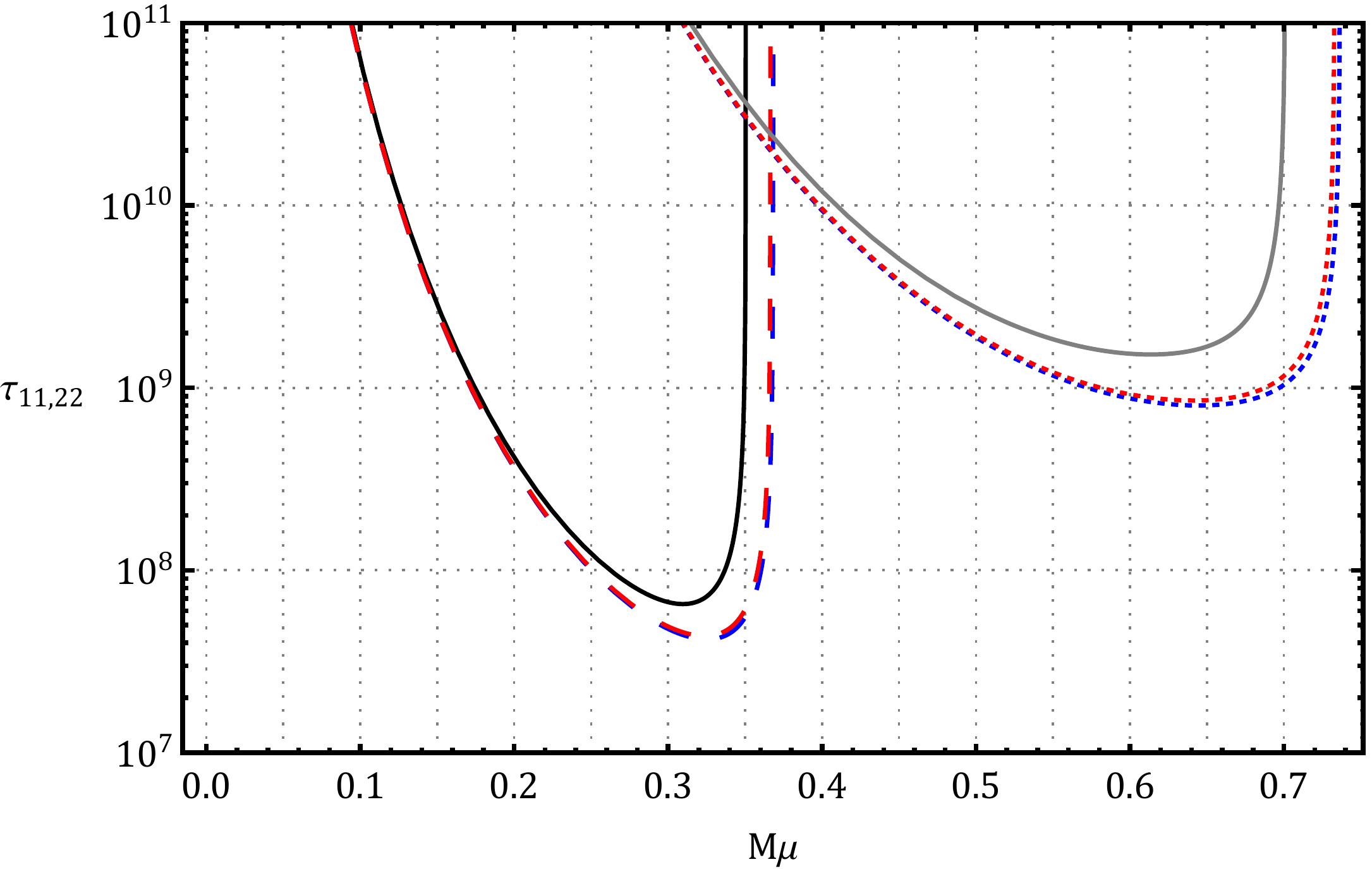}
\end{tabular}
\caption{The timescale of SBS for the fundamental 
	$(n=0)$, $ l = 1=m$ and $l=2=m$ modes (from left to middle) as a function of the gravitational coupling $M\mu$ and for the regular rotating SV black hole with the dimensionless spin parameter $\chi=0.94$. The solid curve represents the standard Kerr black hole, while the blue-dashed and red-dashed curves drawn for the cases  related to the dimensionless regularization parameter $\epsilon=0.02$, with considering the idea that the event horizon radius $r_+$ come from Eqs. (\ref{p}) and (\ref{k}), respectively. The left and the middle panels are integrated in the right panel so that the curves: black-solid, blue-dashed and red-dashed address $M\mu-\tau_{11}$, while gray-solid, blue-dotted, and red-dotted are for $M\mu-\tau_{22}$.}
	\label{Tim}
\end{figure*}

\begin{table*}[ht!]
	\begin{center}	
		{\hfill
			\hbox{
				\begin{tabular}{|c|c|c|c|c|}
					\hline
					$\epsilon$&$M\mu^{*}$&$\tau_{min}^{*}$&$M\mu^{**}$&$\tau_{min}^{**}$\\
					\hline
					$0$&$0.3096$&$6.52\times10^7$ &$0.3096$&$6.52\times10^7$ \\  \hline
					$0.01$&$0.3168$&$5.30\times10^7$ &$0.3165$&$5.35\times10^7$\\  \hline
					$0.02$&$0.3253$&$4.18\times10^7$ &$0.3237$&$4.36\times10^7$ \\ \hline
					$0.03$&$0.3355$&$3.16\times10^7$ &$0.3313$&$3.54\times10^7$ \\ \hline
					$0.04$ &$0.3487$& $2.23\times10^7$ &$0.3392$&$2.86\times10^7$ \\ \hline
					$0.05$&$0.3681$&$1.37\times10^7$ &$0.3475$&$2.30\times10^7$ \\  \hline
				\end{tabular}
			}
			\hfill
			\hbox{
				\begin{tabular}{|c|c|c|c|c|}
					\hline
					$\epsilon$&$M\mu^{*}$&$\tau_{min}^{*}$&$M\mu^{**}$&$\tau_{min}^{**}$\\
					\hline
					$0$&$0.6134$&$1.5238\times10^9$ &$0.6134$&$1.5238\times10^9$ \\  \hline
					$0.01$&$0.6277$&$1.1277\times10^9$ &$0.6271$&$1.1433\times10^9$\\  \hline
					$0.02$&$0.6445$&$8.0070\times10^8$ &$0.6414$&$8.5231\times10^8$ \\ \hline
					$0.03$&$0.6648$&$5.3536\times10^8$ &$0.6564$&$6.3107\times10^8$ \\ \hline
					$0.04$ &$0.6908$& $3.2468\times10^8$ &$0.6721$&$4.6394\times10^8$ \\ \hline
					$0.05$&$0.7293$&$1.6067\times10^8$ &$0.6886$&$3.3854\times10^8$ \\  \hline
				\end{tabular}
			}
			\hfill
		}
		\caption{Numerical values of $\tau_{min}$ in terms of $\epsilon$ for modes: $l=1=m$ (left table) and $l=2=m$  (right table). Symbols * and ** denote to the approximation solutions (\ref{p}), and (\ref{k}), respectively. }
		\label{Nu2}
	\end{center}
\end{table*}

\begin{figure*}[ht!]
\begin{tabular}{c}
%\centering
\includegraphics[width=0.85\columnwidth]{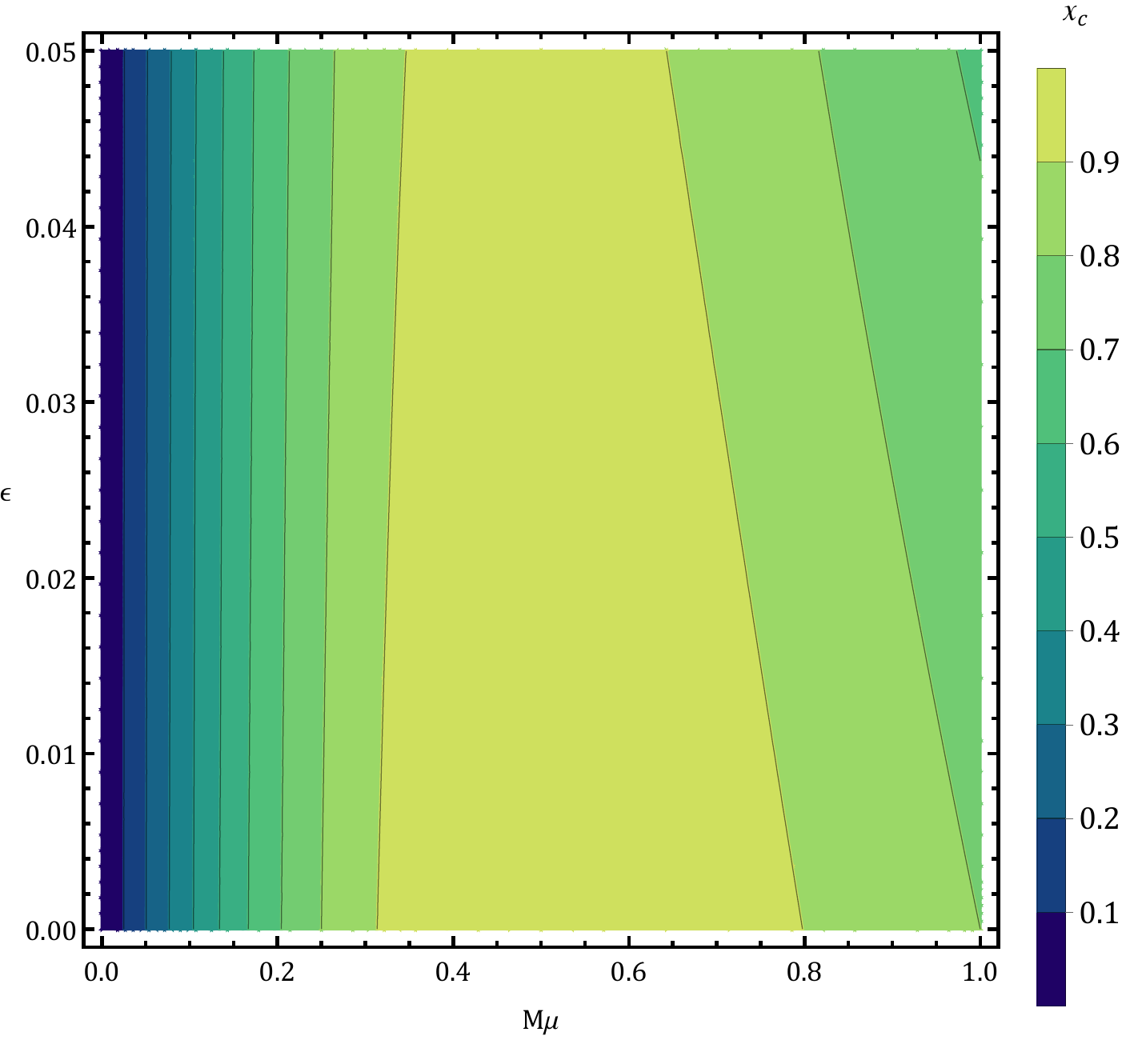}~~~~~~~~~
\includegraphics[width=0.85\columnwidth]{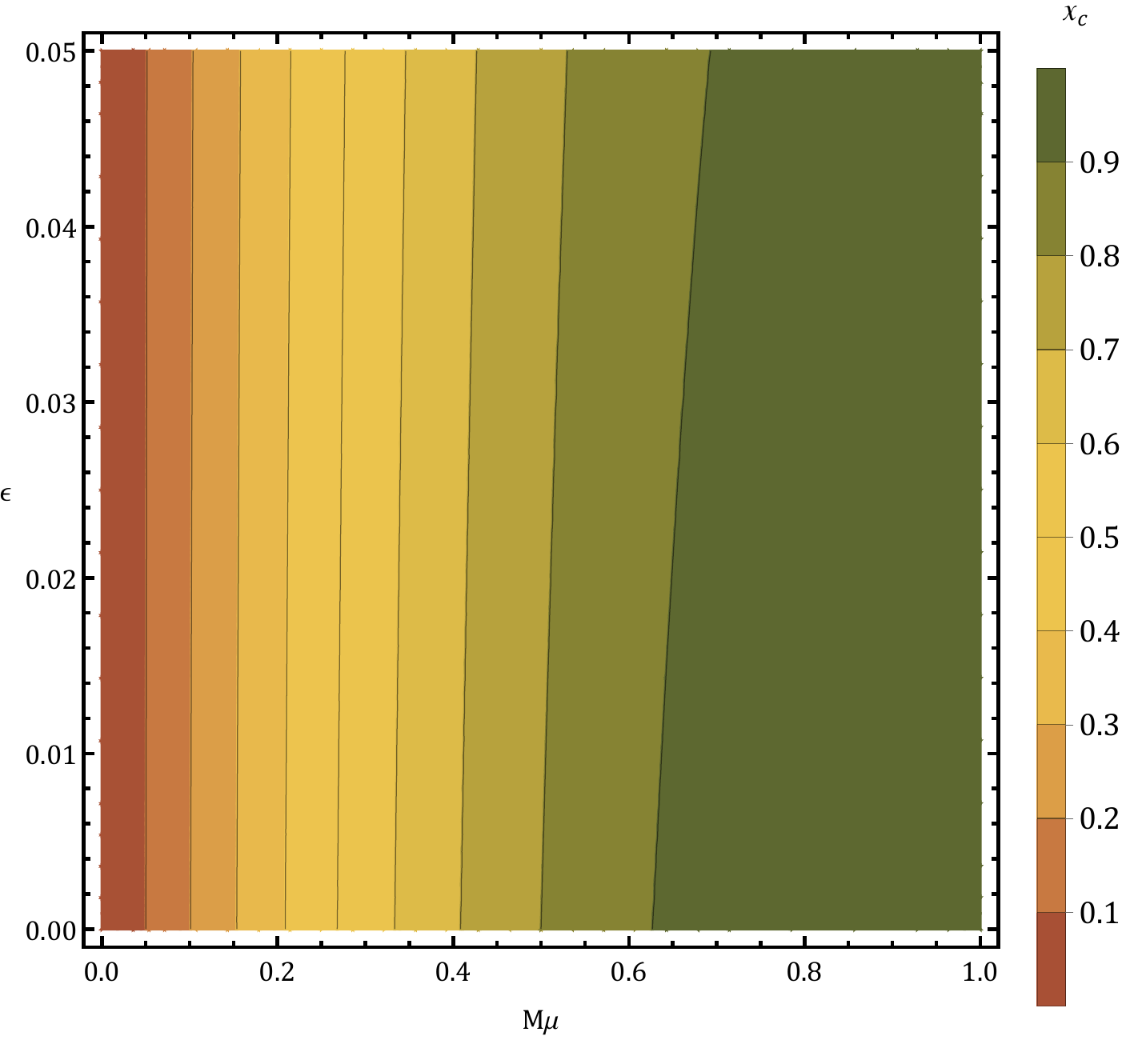}
\end{tabular}
	\caption{A contour plot of the parameter space $M\mu-\epsilon$, for $l = 1=m$, and $l=2=m$ modes (from left to right) showing the critical dimensionless spin parameter $\chi_c$ in which the superradiant instability stops. }
	\label{Con1}
\end{figure*}

\begin{figure*}[ht!]
	\begin{tabular}{c}
		%	\centering
		\includegraphics[width=0.8\columnwidth]{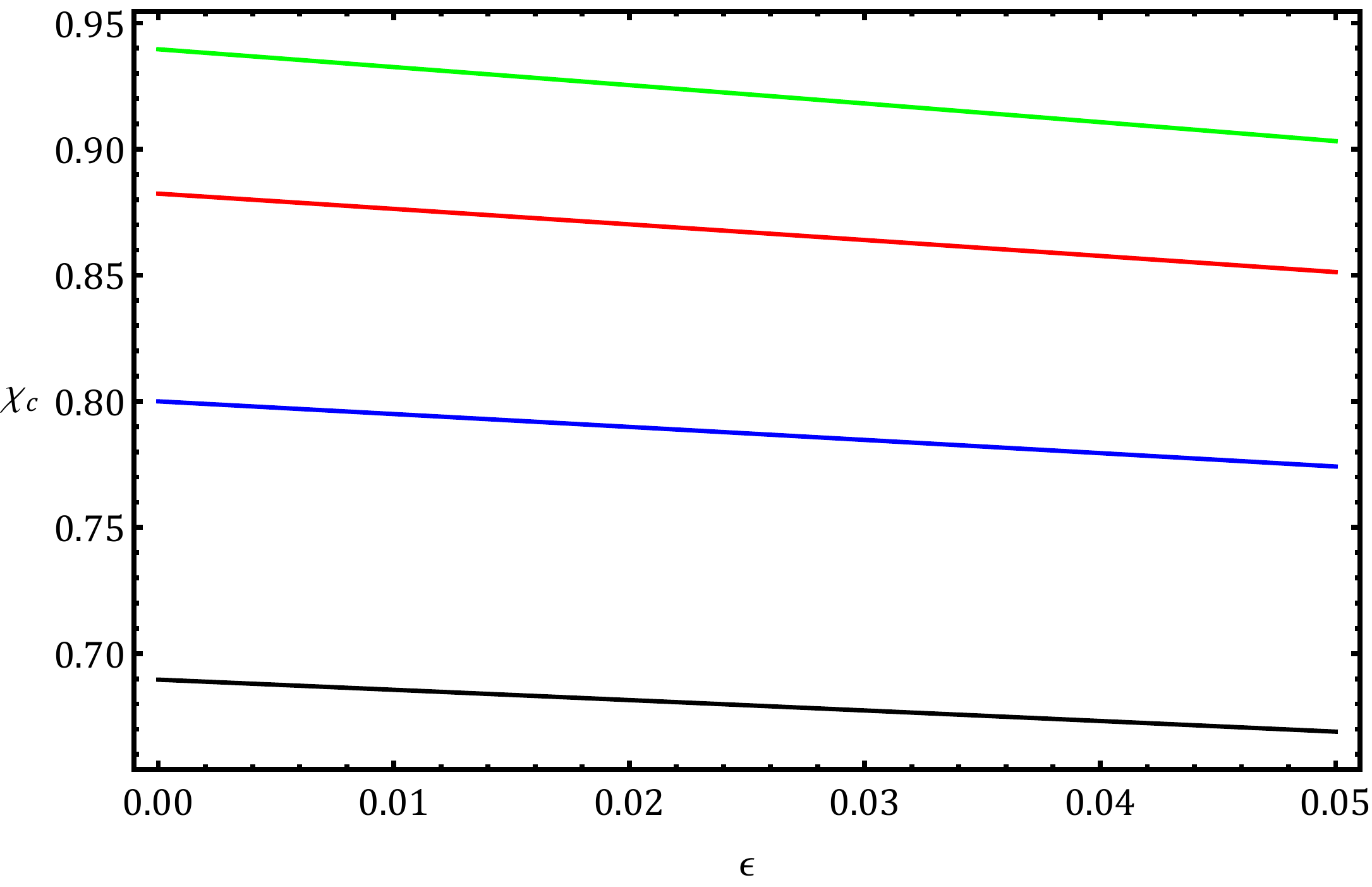}~~~~~~~~~
		\includegraphics[width=0.8\columnwidth]{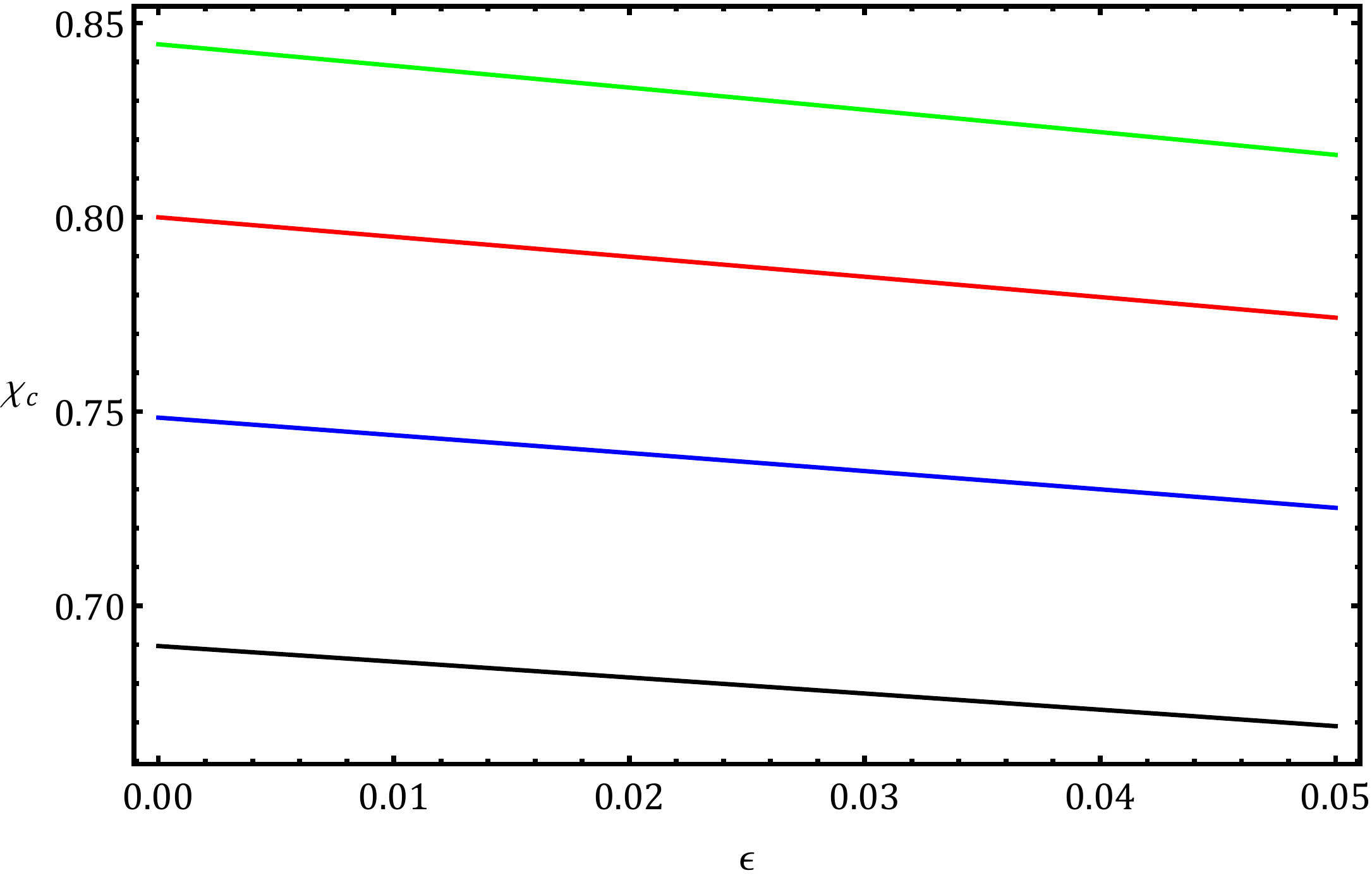}
	\end{tabular}
	\caption{The behavior of the critical value of the dimensionless spin parameter $\chi_c$ in terms of the dimensionless regularization scale $\epsilon$ for different values of the gravitational coupling $M\mu$. In the left panel, by taking the mode $l=1=m$, we set values: $M\mu=0.2, 0.25, 0.3, 0.35$ for black until green, while in the right panel used the values: $M\mu=0.4, 0.45, 0.5, 0.55$, for the mode $l=2=m$.}
	\label{A}
\end{figure*}

\begin{figure}[t]
	\begin{tabular}{c}
		\includegraphics[width=0.9\columnwidth]{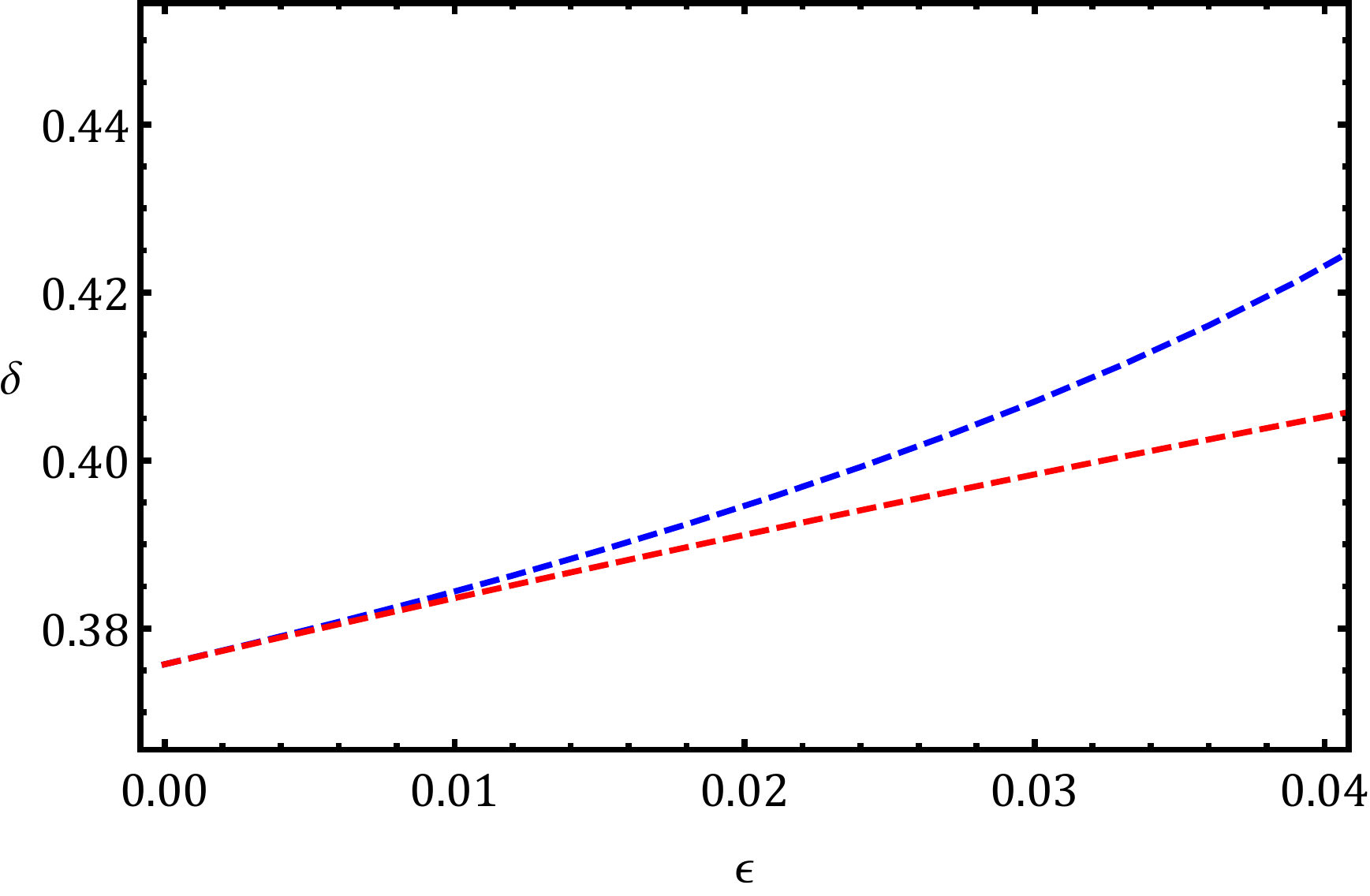}
	\end{tabular}
	\caption{The behavior of ergo-size $\delta$ in terms of the dimensionless regularization scale $\epsilon$ for both approximations used for derivation of the locations of event horizon and ergosphere. For blue-dashed curve we set Eqs. (\ref{p}) and (\ref{erg1}), while for red-dashed, Eqs. (\ref{k}) and (\ref{erg2}). Although, the generality of the behavior of these curves for any value of the dimensionless spin parameter is the same; here, we set $\chi=0.94$, as before.}
	\label{size}
\end{figure}

\begin{figure*}[ht!]
	\begin{tabular}{c}
		%	\centering
		\includegraphics[width=0.8\columnwidth]{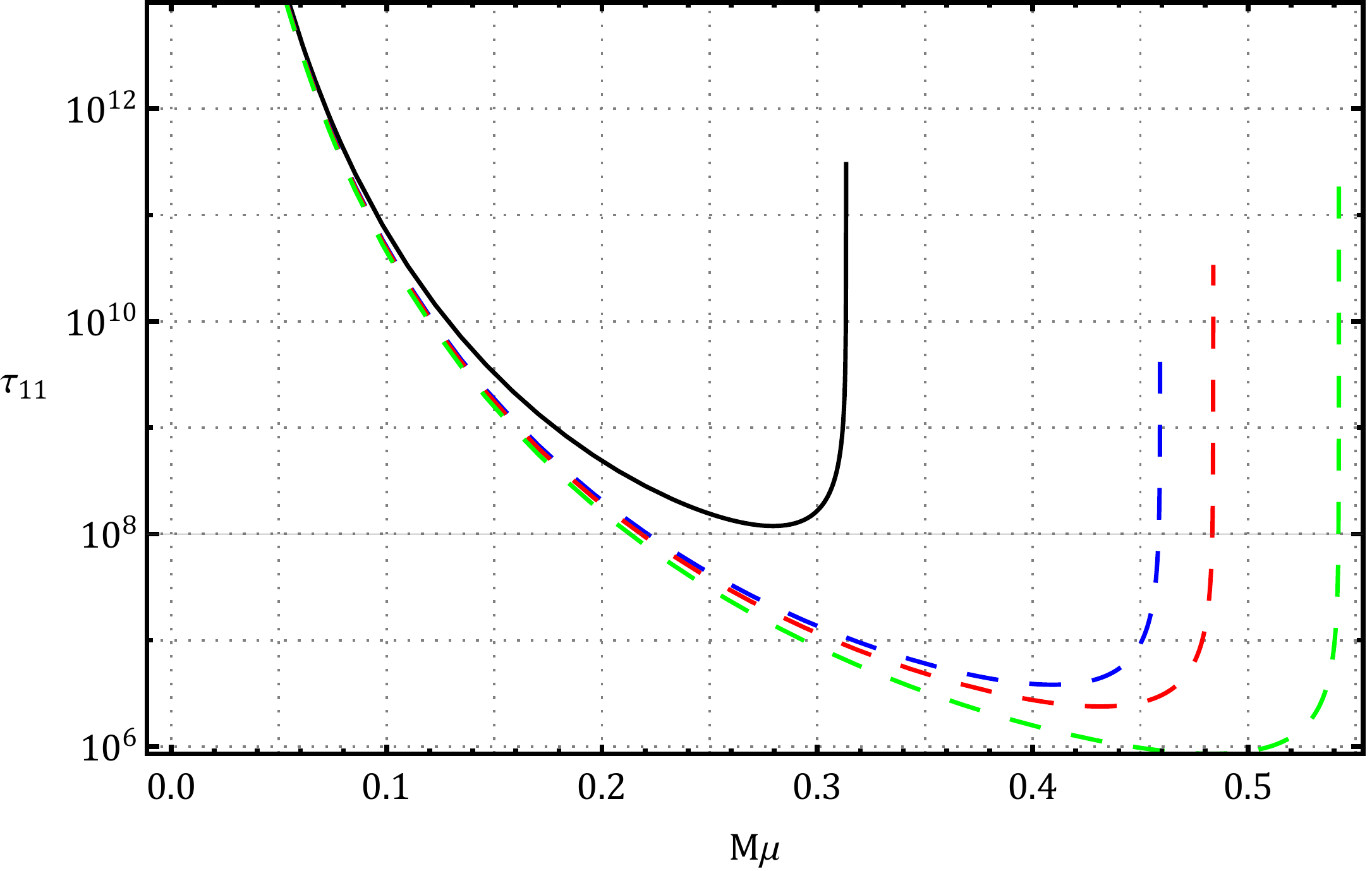}~~~~~~~~~
		\includegraphics[width=0.8\columnwidth]{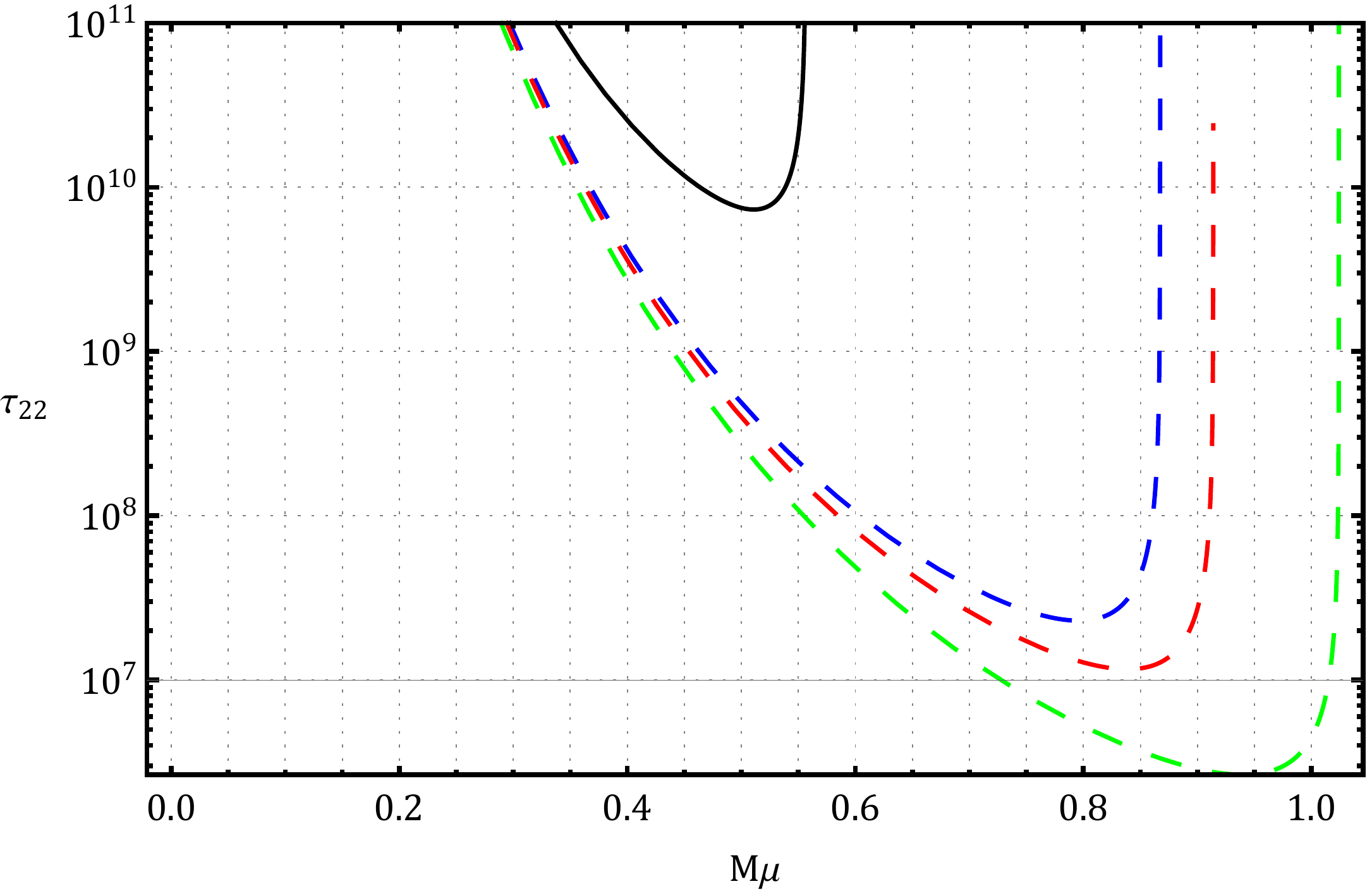}
	\end{tabular}
	\caption{Same as Fig. \ref{Tim}, this time for the non-perturbative values of dimensional regularization parameter $\epsilon$. By taking a rotating black hole with dimensionless spin parameter $\chi=0.85$, we set here numerical values: $0,~0.2,~0.4$ and $0.6$ (from the black-solid to green-dashed curves, respectively) for $\epsilon$. }
	\label{epsilon}
\end{figure*}

By adopting the boundary conditions \eqref{eq:so}, namely an ingoing wave and an exponentially decaying respectively at the event horizon, and at infinity, one will solve the KG equation semi-analytically  \eqref{eq:kg}. Actually, the solution derived in the light of the mentioned boundary conditions can address QBS with the complex frequency $\omega=\omega_R+i\omega_I$. If $\omega_I>0$, then, the amplitude of the perturbation (here the scalar field) grows exponentially with time and subsequently the background turns unstable. Otherwise ($\omega_I<0$), the amplitude of the perturbation dies off exponentially and disappears. The former case has phenomenological worth, in the sense that its inverse reveals the timescale of SCF, as discussed already in the Introduction.

The solutions acquired from the KG equation  \eqref{eq:kg}, within the validity range of the two approximations, $M\mu\ll1$ and $M\omega\ll1$, nicely are equivalent to a Hydrogenic spectrum \cite{Detweiler:1980uk} (see also \cite{Dolan:2007mj}) with a complex frequency   
\begin{align}
&M\omega_R=M\mu\bigg(1-\frac{1}{2}\big(\frac{M\mu}{l+1}\big)^2\bigg)~, \label{r}
\\
&M\omega_I=(\mu M)^{4 l+5}\big( m\chi-2 \mu r_{+}\big) \label{eqi}
\frac{2^{4 l+2}(2 l+1+n) !}{(l+1+n)^{2 l+4} n !}\nonumber \\
& \times\bigg(\frac{l !}{(2 l) !(2 l+1) !}\bigg)^{2} 
 \prod_{k=1}^{l}\bigg(k^{2}\big(1-\chi^2 \big)+\big( m\chi-2 \mu r_{+}\big)^{2}\bigg), 
\end{align}
% where $\chi=a/M$ is the dimensionless spin parameter.
It is clear in case of $\omega_R\approx \mu<m \Omega_+=\frac{ma}{2Mr_+}$, for $m>0$ (modes co-rotating with the black hole), these modes become unstable ($\omega_I>0$) with an instability timescale given by the e-folding time $\tau_{lm}=1/\omega_I$. The gravitational coupling $M\mu$, the spin of the black hole $a$, quantum numbers $(l,m)$, and the location of the event horizon $r_+$ are parameters involved in determining the instability timescale. By taking the fundamental modes with overtone number $n=0$, which address eigenfunctions with zero nodes, for the rotating SV black hole perturbed by the massive scalar field, the results are displayed in the Fig. \ref{Tim}, plots $M\mu-\tau_{11}$ and $M\mu-\tau_{22}$.  
%Note, from now on, in figures we address the regularization length scale via a dimensionless regularization parameter $\epsilon\equiv\frac{\ell}{M}$.
The main reason that the modes $l=m$ are favored commonly, originates from the astrophysical phenomenology in the sense that the timescale of these modes always is shorter than the modes $m<l$.  Right panel in the Fig. \ref{Tim}, confirms this fact that the fastest growth happens for the SBS $l=1=m$. Given that SCF, due to superradiant instability, stops in the saturation points $\omega_R=m\Omega_+$ and $\omega_I \rightarrow 0$, hence the inverse of the maximum of $\omega_I>0$, represents SCF timescale. The minimum of these curves show the values of the gravitational coupling $M\mu$ in which the superradiant instability is the most efficient.
To show the effect of $\epsilon$ in interplay with $M\mu$ on the timescale of SCF as clear as possible, in terms of different perturbative values of $\epsilon$, additionally, we released in Table. \ref{Nu2} some numerical values of $\tau_{min}$ and $M\mu$ in which the superradiant instability is the most efficient. To provide these values, we indeed used the approximated solutions (\ref{p}), and (\ref{k}) to solve equation $\frac{d\tau}{d(M\mu)}=0$. As a consistency check, for the case of $\epsilon=0.02$ one can see a perfect matching between $\tau_{min}$ and $M\mu$ reported in Table. \ref{Nu2}, with the relevant curves in Fig. \ref{Tim}.
The curves in Fig. \ref{Tim} along with the numerical values in Table. \ref{Nu2} give us two intriguing messages about the role of the rotating regular SV black hole. First, in the presence of
perturbative value of dimensionless regularization parameter $\epsilon$ which , in essence, comes from regularization length scale $\ell$; the instability's e-folding time reduces almost up to one order of magnitude, $\mathcal{O}(10)$. Second, embedding the regularization length scale $\ell$ into standard Kerr causes the value of the gravitational coupling $M\mu$ in which the superradiant instability turns efficient, to shift to slightly larger values. One may interpret these outcomes to mean that if the origination of the regularization length scale $\ell$ comes from the Planck length (as has already been noted, one of the domains in which GTR is not valid), it leaves imprints separable from the Kerr. It is interesting potentially, because it can be considered as a smoking gun which may carry some signals from quantum gravity. However, this interpretation may seem crude, and needs to be discussed in more details. In this regard, let us re-express dimensionless regularization parameter $\epsilon$ in the SI unit as follows
\begin{equation}
\epsilon=\frac{\ell}{M}=\frac{l_p c^2}{M_\odot G_N} (\frac{\ell}{l_p})(\frac{M_\odot}{M})\sim10^{-38}~(\frac{\ell}{l_p})(\frac{M_\odot}{M})~,
\end{equation} where $c,~G_N$ and $M_\odot$ are speed of light, gravitational constant and Sun's mass, respectively. Now, by fixing the value of $\ell$ around Planck length $l_p$ to achieve $\epsilon\sim 10^{-2}$, one is required to choose $M\sim 10^{-36}$ $M_\odot$ which does not lie within physically/astrophysically viable range. This shows within a realistic framework the above-mentioned interpretation is not acceptable. Although the story does not end here and there are two other domains for $\ell$ (see the categories quoted from Ref. \cite{Simpson:2021dyo} and listed already) which permit us to set values beyond Planck length. For instance, fixing $\ell=10^{40} l_p$ to obtain $\epsilon\sim 10^{-2}$, we yield $M\sim 10^{4}$ $M_\odot$ which is acceptable astrophysically.

By reading $\omega_I$ from Eq. (\ref{eqi}), and taking positive modes $m>0$, for instance by putting Eq. (\ref{p}) as the event horizon radius \footnote{As noted before, in the case of setting values small enough for $\epsilon$, it has no difference which equation addresses the location of the event horizon, (\ref{p}) or (\ref{k}). }; then, one finds the following critical value for $\chi_c$ to saturate the superradiant condition
\begin{equation}\label{cri}
\chi_c\equiv\dfrac{2M\mu\bigg(m+\sqrt{m^2(1-2\epsilon)-8M^2\mu^2\epsilon}\bigg)}{m^2+4M^2\mu^2}~.
\end{equation}
More precisely, beyond the critical value of the dimensionless spin parameter ($\chi>\chi_c$), we would have superradiant instability. Contour plot drawn in Fig. \ref{Con1}, displays the critical values of $\chi_c$ for saturating the superradiant condition, in interplay with the gravitational coupling $M\mu$ as well as regularization length scale $\ell$. Eq. (\ref{eqi}), indicates clearly the more the difference between $\chi-\chi_c$ i.e., the more the black hole loses rotational energy, the faster the scalar cloud forms. This is exactly the same thing that occurs due to embedding a regularization length scale $\ell$ into the geometry of the Kerr black hole. Actually, by turning on $\ell$ in the Kerr geometry, $\chi_c$ drops with the increment of the suppression parameter's value, as it is evident from Fig. \ref{A}. Namely, in the underlying non-singular background, the energy loss via the superradiant instability stops in lower spins, showing more spin-down relative to its singular counterpart. In other words, $\ell$ causes more rotational energy to be transferred to the scalar cloud, which accelerates its formation.
Indeed, this occurs because embedding a natural regularization length scale $\ell$ into the black hole's geometry causes the superradiant instability to become stronger. Interestingly, it also can be confirmed via the correlation between the strength of the superradiant instability and the measure of the size of the ergoregion, as proposed in \cite{Herdeiro:2014jaa}. There, by introducing a dimensionless criterion $\delta\equiv\frac{A_{erg}-A_+}{16\pi M^2}$, so called ergo-size, indicating the 
difference between the area of the ergo-surface $A_{erg}$ and the area of the event horizon $A_+$; it was proposed the idea arguing that the bigger the $\delta$, the stronger the strength of superradiant instability and vice versa. By taking both approximations into account i.e., Eqs. (\ref{p}), (\ref{k}) for the location of the event horizon and Eqs. (\ref{erg1}), (\ref{erg2}) for the location of ergosphere; we are able to numerically plot $\delta$ in terms of $\epsilon$ in Fig. \ref{size}, to reveal the role of the regularization length scale on the size of ergoregion region around a Kerr black hole. It is clearly shown by adding $\ell$ into standard Kerr geometry, it results in the increment of the ergo-size and subsequently the enhancement of the strength of superradiant instability. Additionally, the increase in $\epsilon$, gives rise to a difference between the curves, as noted earlier. 
This, in it essence, is due to the deviation from a small enough choice of $\epsilon$, as discussed before.

\subsection{ $M\mu-\tau_{lm}$ for non-perturbative values of $\epsilon$}
Here, we are interested in briefly addressing the effect of non-perturbative values of dimensionless regularization parameter $\epsilon$ (i.e., close to $1$) on $\tau_{lm}$. As one can see, due to the approximate solutions (\ref{p})--(\ref{erg2}) obtained from the perturbative procedure, throughout the analysis governed above, we were required to set adequately small values for $\epsilon$. Despite that the modification included in the SV-metric makes it acutely difficult to derive exact solutions, we can alternatively treat it numerically. By adopting such a procedure to derive the location of event horizon and embedding it into Eq. (\ref{eqi}), we drawn plots $M\mu-\tau_{11}$ and $M\mu-\tau_{22}$ in Fig. \ref{epsilon}. 
Contrary to what we perceive from Fig. \ref{Tim}, here $\tau_{min}$ reduces considerably compared to the Kerr case so that as $\epsilon$ increases, the difference becomes more significant. More precisely, taking non-perturbative values of $\epsilon$ into account can lead the lifetime of SCF to reduce more than one order of magnitude which astrophysically is of high value. Particularly, by setting $\epsilon=0.6$ we see with clarity from Fig. \ref{epsilon} that $\tau_{min}$ for modes: $l=1=m$ and $l=2=m$ relative to $\epsilon=0$ is reduced up to two and six orders of magnitude, respectively. Similar to Fig. \ref{Tim}; here, in the presence of $\epsilon$, the value of the gravitational coupling $M\mu$ in which the superradiant instability turns efficient, is shifting to larger values as well; but this time more remarkable, since we set non-perturbative values for $\epsilon$.
%%%%%%%%%%%%%%%%%%%%%%%%%%%%%%%%%%%%%%%%%%%%%%%%%
\section{Conclusion}\label{cons}
It is well known the Kerr spacetime, in case of perturbating by the massive scalar fields, due to black hole bomb phenomena, is prone to superradiant instability. In this way,  the black hole mimics a gravitational atom, due to the quasi-stationary gravitational bound states created by the scalar clouds arising from exponential amplification of an initial scalar perturbation around the black hole. Actually, we deal with Hydrogen atom-like bound states between the black hole and the scalar cloud. The bosonic scalar clouds around the rotating black hole are of phenomenological importance, since via gravitational wave emission, they can be used to probe new physics in the form of ultra-light particles. The key concern in this scenario is related to the required growth time for superradiant bound states i.e., the lifetime of time-growing quasi-bound states. More precisely, the shorter the timescale of the scalar cloud formation, the more the astrophysical importance.

In this paper, we have addressed this question: even though the singularity is hidden within the event horizon, whether bypassing the central singularity of the Kerr black hole affect the timescale of the scalar cloud formation around the black hole or not? Bypassing the singularity, in essence, is an inevitable requirement in the General Theory of Relativity which is expected to be successfully performed by including the quantum gravity considerations. In the absence of a full theory of quantum gravity, regular models as frameworks to effectively remove curvature singularity in the Kerr black hole have received much attention. One of these regular models which was recently proposed by Simpson and Visser is a rotating hollow-like regular black hole with an asymptotically Minkowski core, including some desirable features which are missing in its standard counterparts, and made it an astrophysically viable black hole. This has led us to consider this novel regular model as a merit framework for addressing the above question.

In the first step, we have investigated the superradiant instability of the regular rotating Simpson-Visser black hole, which is perturbed by a massive scalar field. In this metric, there is a parameter, so-called regularization length scale $\ell$, in addition to mass $M$ and spin $a$ which deviates it from the singular Kerr metric. Our instability analysis in the form of black hole bomb phenomena shows the standard superradiant instability regime does not get affected by $\ell$.
 In the next step, by translating $\ell$ into a dimensionless parameter $\epsilon=\ell/M$ we have investigated the effect of regularization on the lifetime of quasi-bound states formed within the relevant superradiant instability regime; which are known as the superradiant-bound states as well.
 This is well-motivated, since it lets us know about the role of regularization of the central singularity in Kerr's black hole on the timescale of the scalar cloud formation around the black hole which, in essence, comes from superradiant bound states' formation. By taking the hydrogen-like approximation, we shown a scalar cloud around the regularized Simpson-Visser black hole, forms faster than the standard Kerr case. 
More detailed, adding the regularization length scale into the Kerr metric causes the timescale of the formation of a scalar cloud around a rotating black hole to become shorter (from one ($\mathcal{O}(10)$) to a few orders of magnitude, depending on the values set for $\epsilon$). Likewise we argued this positive outcome can be applied to astrophysical black holes when the regularization length scale embedded in the Kerr spacetime exceeds the Planck length. It has no contradiction with the framework proposed by Simpson and Visser because one is free to choose the scale of the parameter $\ell$.
We have shown the enhancement of the superradiant instability's strength due to the increase in the ergo-size, is the origin of the acceleration of the formation of scalar clouds around the regular Kerr black hole at hand, compared to the standard one.

From the perspective of phenomenology in the framework of astrophysics, this may be interesting since the scalar cloud, after its formation, is able to play the role of a continuum source of gravitational waves via energy dissipation.
By and large, the important message of this study is that demanding a natural and inevitable requirement such as bypassing the central singularity of the standard Kerr metric, results in a shorter timescale of the scalar cloud formation, which improves the detection prospects of new physics (particularly ultra-light particles).

\vspace{0.3cm}
{\bf Acknowledgments:}
M. Kh appreciates Carlos Herdeiro for reading and constructive  comments on the manuscript and Javad T. Firouzjaee for clarification on some questions. We thank the anonymous referee for comments that led to improvements in this manuscript.
%%%%%%%%%%%%%%%%%%%%%%%%%%%%%%%%%%%%%%%%%%%%%%%%%%%%


\begin{thebibliography}{10}

\bibitem{Regge:1957td}
T.~Regge and J.~A.~Wheeler,
%``Stability of a Schwarzschild singularity,''
Phys. Rev. \textbf{108} (1957), 1063-1069
%doi:10.1103/PhysRev.108.1063

\bibitem{Vishveshwara:1970cc}
C.~V.~Vishveshwara,
%``Stability of the schwarzschild metric,''
Phys. Rev. D \textbf{1} (1970), 2870-2879
%doi:10.1103/PhysRevD.1.2870

\bibitem{Zerilli:1970wzz}
F.~J.~Zerilli,
%``Gravitational field of a particle falling in a schwarzschild geometry analyzed in tensor harmonics,''
Phys. Rev. D \textbf{2} (1970), 2141-2160
%doi:10.1103/PhysRevD.2.2141


\bibitem{Teukolsky:1972my}
S.~A.~Teukolsky,
%``Rotating black holes - separable wave equations for gravitational and electromagnetic perturbations,''
Phys. Rev. Lett. \textbf{29} (1972), 1114-1118
%doi:10.1103/PhysRevLett.29.1114

\bibitem{Press:1973zz}
W.~H.~Press and S.~A.~Teukolsky,
%``Perturbations of a Rotating Black Hole. II. Dynamical Stability of the Kerr Metric,''
Astrophys. J. \textbf{185} (1973), 649-674
%doi:10.1086/152445

\bibitem{Zel:1971}
Ya. B. Zel'dovich, JETP Lett. \textbf{14} (1971) 180

\bibitem{Zel:1972}
Ya. B. Zel'dovich, Sov. Phys. JETP \textbf{35}  (1972) 1085

\bibitem{Starobinsky:1973aij}
A.~A.~Starobinsky,
%``Amplification of waves reflected from a rotating ''black hole''.,''
Sov. Phys. JETP \textbf{37} (1973) no.1, 28-32

\bibitem{Penrose:1969pc}
R.~Penrose,
%``Gravitational collapse: The role of general relativity,''
Gen.\ Rel.\ Grav.\  {\bf 34} (2002) 1141
%doi:10.1023/A:1016578408204

\bibitem{Comisso:2020ykg}
L.~Comisso and F.~A.~Asenjo,
%``Magnetic Reconnection as a Mechanism for Energy Extraction from Rotating Black Holes,''
Phys.\ Rev.\ D {\bf 103} (2021) no.2,  023014
%doi:10.1103/PhysRevD.103.023014
[arXiv:2012.00879 [astro-ph.HE]].

\bibitem{Khodadi:2022dff}
M.~Khodadi,
%``Magnetic reconnection and energy extraction from a spinning black hole with broken Lorentz symmetry,''
Phys. Rev. D \textbf{105} (2022) no.2, 023025
%doi:10.1103/PhysRevD.105.023025
[arXiv:2201.02765 [gr-qc]].

\bibitem{Hawking:1975vcx}
S.~W.~Hawking,
%``Particle Creation by Black Holes,''
Commun. Math. Phys. \textbf{43} (1975), 199-220
[erratum: Commun. Math. Phys. \textbf{46} (1976), 206]
%doi:10.1007/BF02345020


\bibitem{Pani:2011gy}
P.~Pani, C.~F.~B.~Macedo, L.~C.~B.~Crispino and V.~Cardoso,
%``Slowly rotating black holes in alternative theories of gravity,''
Phys.\ Rev.\ D {\bf 84} (2011) 087501
	%doi:10.1103/PhysRevD.84.087501
	[arXiv:1109.3996 [gr-qc]].

\bibitem{Kleihaus:2011tg}
B.~Kleihaus, J.~Kunz and E.~Radu,
%``Rotating Black Holes in Dilatonic Einstein-Gauss-Bonnet Theory,''
Phys.\ Rev.\ Lett.\  {\bf 106} (2011) 151104
	%doi:10.1103/PhysRevLett.106.151104
	[arXiv:1101.2868 [gr-qc]].

\bibitem{Delsate:2018ome}
T.~Delsate, C.~Herdeiro and E.~Radu,
%``Non-perturbative spinning black holes in dynamical Chern–Simons gravity,''
Phys.\ Lett.\ B {\bf 787} (2018) 8
	%doi:10.1016/j.physletb.2018.09.060
	[arXiv:1806.06700 [gr-qc]].


\bibitem{Cardoso:2013opa}
V.~Cardoso, I.~P.~Carucci, P.~Pani and T.~P.~Sotiriou,
%``Matter around Kerr black holes in scalar-tensor theories: scalarization and superradiant instability,''
Phys.\ Rev.\ D {\bf 88} (2013) 044056
	%doi:10.1103/PhysRevD.88.044056
	[arXiv:1305.6936 [gr-qc]].

\bibitem{Cardoso:2013fwa}
V.~Cardoso, I.~P.~Carucci, P.~Pani and T.~P.~Sotiriou,
%``Black holes with surrounding matter in scalar-tensor theories,''
Phys.\ Rev.\ Lett.\  {\bf 111} (2013) 111101
	%doi:10.1103/PhysRevLett.111.111101
	[arXiv:1308.6587 [gr-qc]].

\bibitem{Aliev:2014aba}
A.~N.~Aliev,
%``Superradiance and black hole bomb in five-dimensional minimal ungauged supergravity,''
JCAP {\bf 1411} (2014) 029
%  doi:10.1088/1475-7516/2014/11/029
	[arXiv:1408.4269 [hep-th]].

\bibitem{Zhang:2014kna}
C.~Y.~Zhang, S.~J.~Zhang and B.~Wang,
%``Superradiant instability of Kerr-de Sitter black holes in scalar-tensor theory,''
JHEP {\bf 1408} (2014) 011
	%doi:10.1007/JHEP08(2014)011
	[arXiv:1405.3811 [hep-th]].

\bibitem{Fierro:2017fky}
O.~Fierro, N.~Grandi and J.~Oliva,
%``Superradiance of charged black holes in Einstein–Gauss–Bonnet gravity,''
Class.\ Quant.\ Grav.\  {\bf 35} (2018) no.10,  105007
	%doi:10.1088/1361-6382/aab3f6
	[arXiv:1708.06037 [hep-th]].

%\bibitem{Wondrak:2018fza}
%M.~F.~Wondrak, P.~Nicolini and J.~W.~Moffat,
%``Superradiance in Modified Gravity (MOG),''
%JCAP {\bf 1812} (2018) 021
%doi:10.1088/1475-7516/2018/12/021
%[arXiv:1809.07509 [gr-qc]].

\bibitem{Kolyvaris:2018zxl}
T.~Kolyvaris, M.~Koukouvaou, A.~Machattou and E.~Papantonopoulos,
%``Superradiant instabilities in scalar-tensor Horndeski theory,''
Phys.\ Rev.\ D {\bf 98} (2018) no.2,  024045
	%doi:10.1103/PhysRevD.98.024045
	[arXiv:1806.11110 [gr-qc]].

\bibitem{Ficarra:2018rfu}
G.~Ficarra, P.~Pani and H.~Witek,
%``Impact of multiple modes on the black-hole superradiant instability,''
Phys. Rev. D \textbf{99} (2019) no.10, 104019
%doi:10.1103/PhysRevD.99.104019
[arXiv:1812.02758 [gr-qc]].

\bibitem{Khodadi:2020cht}
M.~Khodadi, A.~Talebian and H.~Firouzjahi,
%``Black Hole Superradiance in f(R) Gravities,''
arXiv:2002.10496 [gr-qc].

\bibitem{Zhang:2020sjh}
C.~Y.~Zhang, S.~J.~Zhang, P.~C.~Li and M.~Guo,
%``Superradiance and stability of the regularized 4D charged Einstein-Gauss-Bonnet black hole,''
JHEP \textbf{08} (2020), 105
	%doi:10.1007/JHEP08(2020)105
	[arXiv:2004.03141 [gr-qc]].


\bibitem{Khodadi:2021owg}
M.~Khodadi,
%``Black Hole Superradiance in the Presence of Lorentz Symmetry Violation,''
Phys. Rev. D \textbf{103} (2021) no.6, 064051
%doi:10.1103/PhysRevD.103.064051
[arXiv:2103.03611 [gr-qc]].



\bibitem{Mehta:2021pwf}
V.~M.~Mehta, M.~Demirtas, C.~Long, D.~J.~E.~Marsh, L.~McAllister and M.~J.~Stott,
%``Superradiance in string theory,''
JCAP \textbf{07} (2021), 033
	%doi:10.1088/1475-7516/2021/07/033
	[arXiv:2103.06812 [hep-th]].

\bibitem{Jiang:2021whw}
R.~Jiang, R.~H.~Lin and X.~H.~Zhai,
%``Superradiant instability of a Kerr-like black hole in Einstein-bumblebee gravity,''
Phys. Rev. D \textbf{104} (2021) no.12, 124004
%doi:10.1103/PhysRevD.104.124004
[arXiv:2108.04702 [gr-qc]].


\bibitem{Khodadi:2021mct}
M.~Khodadi and R.~Pourkhodabakhshi,
%``Superradiance and stability of Kerr black hole enclosed by anisotropic fluid matter,''
Phys. Lett. B \textbf{823} (2021), 136775
%doi:10.1016/j.physletb.2021.136775
[arXiv:2111.03316 [gr-qc]].

\bibitem{Jha:2021bue}
S.~K.~Jha and A.~Rahaman,
%``Lorentz violation and noncommutative effect on superradiance scattering off Kerr-like black hole and on the shadow of it,''
[arXiv:2111.02817 [gr-qc]].

\bibitem{Jha:2022ewi}
S.~K.~Jha and A.~Rahaman,
%``Kerr\textendash{}Sen-like Lorentz violating black holes and superradiance phenomena,''
Eur. Phys. J. C \textbf{82} (2022) no.5, 411
%doi:10.1140/epjc/s10052-022-10307-y
[arXiv:2203.08099 [gr-qc]].



\bibitem{Press:1972zz}
W.~H.~Press and S.~A.~Teukolsky,
%``Floating Orbits, Superradiant Scattering and the Black-hole Bomb,''
Nature \textbf{238} (1972), 211-212.
%doi:10.1038/238211a0

\bibitem{Cardoso:2004nk}
V.~Cardoso, O.~J.~C.~Dias, J.~P.~S.~Lemos and S.~Yoshida,
%``The Black hole bomb and superradiant instabilities,''
Phys. Rev. D \textbf{70} (2004), 044039
[erratum: Phys. Rev. D \textbf{70} (2004), 049903]
%doi:10.1103/PhysRevD.70.049903
[arXiv:hep-th/0404096 [hep-th]].


\bibitem{Furuhashi:2004jk}
H.~Furuhashi and Y.~Nambu,
%``Instability of massive scalar fields in Kerr-Newman space-time,''
Prog. Theor. Phys. \textbf{112} (2004), 983-995
	%doi:10.1143/PTP.112.983
	[arXiv:gr-qc/0402037 [gr-qc]].


\bibitem{Dolan:2007mj}
S.~R.~Dolan,
%``Instability of the massive Klein-Gordon field on the Kerr spacetime,''
Phys. Rev. D \textbf{76} (2007), 084001
	%doi:10.1103/PhysRevD.76.084001
	[arXiv:0705.2880 [gr-qc]].

\bibitem{Hod:2012zza}
S.~Hod,
%``On the instability regime of the rotating Kerr spacetime to massive scalar perturbations,''
Phys.\ Lett.\ B {\bf 708} (2012) 320
	%  doi:10.1016/j.physletb.2012.01.054
	[arXiv:1205.1872 [gr-qc]].


\bibitem{Dolan:2012yt} 
S.~R.~Dolan,
%``Superradiant instabilities of rotating black holes in the time domain,''
Phys. Rev. D \textbf{87} (2013) no.12, 124026
	%doi:10.1103/PhysRevD.87.124026
	[arXiv:1212.1477 [gr-qc]].


\bibitem{Zhu:2014sya}
Z.~Zhu, S.~J.~Zhang, C.~E.~Pellicer, B.~Wang and E.~Abdalla,
%``Stability of Reissner-Nordstr\"om black hole in de Sitter background under charged scalar perturbation,''
Phys. Rev. D \textbf{90} (2014) no.4, 044042
	%doi:10.1103/PhysRevD.90.044042
	[arXiv:1405.4931 [hep-th]].


\bibitem{Huang:2018qdl}
Y.~Huang, D.~J.~Liu, X.~h.~Zhai and X.~z.~Li,
%``Instability for massive scalar fields in Kerr-Newman spacetime,''
Phys. Rev. D \textbf{98} (2018) no.2, 025021
	%doi:10.1103/PhysRevD.98.025021
	[arXiv:1807.06263 [gr-qc]].


\bibitem{Destounis:2019hca}
K.~Destounis,
%``Superradiant instability of charged scalar fields in higher-dimensional Reissner-Nordström-de Sitter black holes,''
Phys.\ Rev.\ D {\bf 100} (2019) no.4,  044054
	%  doi:10.1103/PhysRevD.100.044054
	[arXiv:1908.06117 [gr-qc]].

\bibitem{Huang:2019xbu}
J.~H.~Huang, W.~X.~Chen, Z.~Y.~Huang and Z.~F.~Mai,
%``Superradiant stability of the Kerr black holes,''
Phys.\ Lett.\ B {\bf 798} (2019) 135026
	% doi:10.1016/j.physletb.2019.135026
	[arXiv:1907.09118 [gr-qc]].


\bibitem{Xu:2020fgq}
J.~H.~Xu, Z.~H.~Zheng, M.~J.~Luo and J.~H.~Huang,
%``Analytic study of superradiant stability of Kerr\textendash{}Newman black holes under charged massive scalar perturbation,''
Eur. Phys. J. C \textbf{81} (2021) no.5, 402
	%doi:10.1140/epjc/s10052-021-09180-y
	[arXiv:2012.13594 [gr-qc]].

\bibitem{Vieira:2021nha}
H.~S.~Vieira, V.~B.~Bezerra and C.~R.~Muniz,
%``(In)stability of the charged massive scalar field on the Kerr-Newman spacetime,''
[arXiv:2107.02562 [gr-qc]].


\bibitem{Cardoso:2006wa}
V.~Cardoso, O.~J.~C.~Dias and S.~Yoshida,
%``Classical instability of Kerr-AdS black holes and the issue of final state,''
Phys.\ Rev.\ D {\bf 74} (2006) 044008
	%doi:10.1103/PhysRevD.74.044008
	[hep-th/0607162].

\bibitem{Li:2012rx}
R.~Li,
%``Superradiant instability of charged massive scalar field in Kerr-Newman-anti-de Sitter black hole,''
Phys.\ Lett.\ B {\bf 714} (2012) 337
	%doi:10.1016/j.physletb.2012.07.015
	[arXiv:1205.3929 [gr-qc]].

\bibitem{Green:2015kur}
S.~R.~Green, S.~Hollands, A.~Ishibashi and R.~M.~Wald,
%``Superradiant instabilities of asymptotically anti-de Sitter black holes,''
Class.\ Quant.\ Grav.\  {\bf 33} (2016) no.12,  125022
%doi:10.1088/0264-9381/33/12/125022
[arXiv:1512.02644 [gr-qc]].
	

\bibitem{Li:2019tns}
R.~Li, Y.~Zhao, T.~Zi and X.~Chen,
%``Superradiance and dynamical evolution of a charged scalar field in an asymptotically anti–de-Sitter dilatonic black hole,''
Phys.\ Rev.\ D {\bf 99} (2019) no.8,  084045.
% doi:10.1103/PhysRevD.99.084045

\bibitem{Rahmani:2020wlq}
A.~Rahmani, M.~Honardoost and H.~R.~Sepangi,
%``Thermal phase transition in $F(R)$-charged $AdS_{4}$-scalar theory,''
Phys. Rev. D \textbf{101} (2020) no.8, 084036
%doi:10.1103/PhysRevD.101.084036
[arXiv:2002.01663 [gr-qc]].

\bibitem{Rahmani:2020vvv}
A.~Rahmani, M.~Khodadi, M.~Honardoost and H.~R.~Sepangi,
%``Instability and no-hair paradigm in d-dimensional charged-AdS black holes,''
Nucl. Phys. B \textbf{960} (2020), 115185
%doi:10.1016/j.nuclphysb.2020.115185
[arXiv:2009.09186 [gr-qc]].


\bibitem{Grain:2007gn}
J.~Grain and A.~Barrau,
%``Quantum bound states around black holes,''
Eur. Phys. J. C \textbf{53} (2008), 641-648
%doi:10.1140/epjc/s10052-007-0494-1
[arXiv:hep-th/0701265 [hep-th]].


\bibitem{Hod:2017gvn}
S.~Hod,
%``Quasi-bound state resonances of charged massive scalar fields in the near-extremal Reissner\textendash{}Nordstr\"om black-hole spacetime,''
Eur. Phys. J. C \textbf{77} (2017) no.5, 351
%doi:10.1140/epjc/s10052-017-4920-8
[arXiv:1705.04726 [hep-th]].

\bibitem{Baryakhtar:2020gao}
M.~Baryakhtar, M.~Galanis, R.~Lasenby and O.~Simon,
%``Black hole superradiance of self-interacting scalar fields,''
Phys. Rev. D \textbf{103} (2021) no.9, 095019
%doi:10.1103/PhysRevD.103.095019
[arXiv:2011.11646 [hep-ph]].


\bibitem{Detweiler:1973zz}
S.~L.~Detweiler and J.~R.~Ipser,
%``Stability of scalar perturbations of a Kerr-metric black hole,''
Astrophys. J. \textbf{185} (1973), 675-683
%doi:10.1086/152446

\bibitem{Hod:2012wmy}
S.~Hod,
%``Stability of the extremal Reissner-Nordstroem black hole to charged scalar perturbations,''
Phys. Lett. B \textbf{713} (2012), 505-508
%doi:10.1016/j.physletb.2012.06.043
[arXiv:1304.6474 [gr-qc]].

\bibitem{Hod:2013nn}
S.~Hod,
%``No-bomb theorem for charged Reissner-Nordstroem black holes,''
Phys. Lett. B \textbf{718} (2013), 1489-1492
%doi:10.1016/j.physletb.2012.12.013


\bibitem{Brito:2015oca}
R.~Brito, V.~Cardoso and P.~Pani,
%``Superradiance: New Frontiers in Black HolePhysics,''
Lect. Notes Phys. \textbf{906} (2015), pp.1-237
%doi:10.1007/978-3-319-19000-6
[arXiv:1501.06570 [gr-qc]].




\bibitem{Sampaio:2014swa}
M.~O.~P.~Sampaio, C.~Herdeiro and M.~Wang,
%``Marginal scalar and Proca clouds around Reissner-Nordstr\"om black holes,''
Phys. Rev. D \textbf{90} (2014) no.6, 064004
%doi:10.1103/PhysRevD.90.064004
[arXiv:1406.3536 [gr-qc]].


\bibitem{Herdeiro:2014goa}
C.~A.~R.~Herdeiro and E.~Radu,
%``Kerr black holes with scalar hair,''
Phys. Rev. Lett. \textbf{112} (2014), 221101
%doi:10.1103/PhysRevLett.112.221101
[arXiv:1403.2757 [gr-qc]].


\bibitem{East:2017ovw}
W.~E.~East and F.~Pretorius,
%``Superradiant Instability and Backreaction of Massive Vector Fields around Kerr Black Holes,''
Phys. Rev. Lett. \textbf{119} (2017) no.4, 041101
%doi:10.1103/PhysRevLett.119.041101
[arXiv:1704.04791 [gr-qc]].

\bibitem{East:2017mrj}
W.~E.~East,
%``Superradiant instability of massive vector fields around spinning black holes in the relativistic regime,''
Phys. Rev. D \textbf{96} (2017) no.2, 024004
%doi:10.1103/PhysRevD.96.024004
[arXiv:1705.01544 [gr-qc]].

\bibitem{Herdeiro:2017phl}
C.~A.~R.~Herdeiro and E.~Radu,
%``Dynamical Formation of Kerr Black Holes with Synchronized Hair: An Analytic Model,''
Phys. Rev. Lett. \textbf{119} (2017) no.26, 261101
%doi:10.1103/PhysRevLett.119.261101
[arXiv:1706.06597 [gr-qc]].


\bibitem{East:2018glu}
W.~E.~East,
%``Massive Boson Superradiant Instability of Black Holes: Nonlinear Growth, Saturation, and Gravitational Radiation,''
Phys. Rev. Lett. \textbf{121} (2018) no.13, 131104
%doi:10.1103/PhysRevLett.121.131104
[arXiv:1807.00043 [gr-qc]].


\bibitem{EventHorizonTelescope:2019dse}
K.~Akiyama \textit{et al.} [Event Horizon Telescope],
%``First M87 Event Horizon Telescope Results. I. The Shadow of the Supermassive Black Hole,''
Astrophys. J. Lett. \textbf{875} (2019), L1
%doi:10.3847/2041-8213/ab0ec7
[arXiv:1906.11238 [astro-ph.GA]].


\bibitem{Creci:2020mfg}
G.~Creci, S.~Vandoren and H.~Witek,
%``Evolution of black hole shadows from superradiance,''
Phys. Rev. D \textbf{101} (2020) no.12, 124051
%doi:10.1103/PhysRevD.101.124051
[arXiv:2004.05178 [gr-qc]].

\bibitem{Roy:2021uye}
R.~Roy, S.~Vagnozzi and L.~Visinelli,
%``Superradiance evolution of black hole shadows revisited,''
Phys. Rev. D \textbf{105} (2022) no.8, 083002
%doi:10.1103/PhysRevD.105.083002
[arXiv:2112.06932 [astro-ph.HE]].


\bibitem{Roy:2019esk}
R.~Roy and U.~A.~Yajnik,
%``Evolution of black hole shadow in the presence of ultralight bosons,''
Phys. Lett. B \textbf{803} (2020), 135284
%doi:10.1016/j.physletb.2020.135284
[arXiv:1906.03190 [gr-qc]].

\bibitem{Cunha:2019ikd}
P.~V.~P.~Cunha, C.~A.~R.~Herdeiro and E.~Radu,
%``EHT constraint on the ultralight scalar hair of the M87 supermassive black hole,''
Universe \textbf{5} (2019) no.12, 220
%doi:10.3390/universe5120220
[arXiv:1909.08039 [gr-qc]].


\bibitem{Yuan:2021ebu}
C.~Yuan, R.~Brito and V.~Cardoso,
%``Probing ultralight dark matter with future ground-based gravitational-wave detectors,''
Phys. Rev. D \textbf{104} (2021) no.4, 044011
%doi:10.1103/PhysRevD.104.044011
[arXiv:2106.00021 [gr-qc]].

\bibitem{Arvanitaki:2014wva}
A.~Arvanitaki, M.~Baryakhtar and X.~Huang,
%``Discovering the QCD Axion with Black Holes and Gravitational Waves,''
Phys. Rev. D \textbf{91} (2015) no.8, 084011
%doi:10.1103/PhysRevD.91.084011
[arXiv:1411.2263 [hep-ph]].


\bibitem{Arvanitaki:2016qwi}
A.~Arvanitaki, M.~Baryakhtar, S.~Dimopoulos, S.~Dubovsky and R.~Lasenby,
%``Black Hole Mergers and the QCD Axion at Advanced LIGO,''
Phys. Rev. D \textbf{95} (2017) no.4, 043001
%doi:10.1103/PhysRevD.95.043001
[arXiv:1604.03958 [hep-ph]].


\bibitem{Ghosh:2018gaw}
S.~Ghosh, E.~Berti, R.~Brito and M.~Richartz,
%``Follow-up signals from superradiant instabilities of black hole merger remnants,''
Phys. Rev. D \textbf{99} (2019) no.10, 104030
%doi:10.1103/PhysRevD.99.104030
[arXiv:1812.01620 [gr-qc]].

\bibitem{Brito:2020lup}
R.~Brito, S.~Grillo and P.~Pani,
%``Black Hole Superradiant Instability from Ultralight Spin-2 Fields,''
Phys. Rev. Lett. \textbf{124} (2020) no.21, 211101
%doi:10.1103/PhysRevLett.124.211101
[arXiv:2002.04055 [gr-qc]].

\bibitem{Baumann:2022pkl}
D.~Baumann, G.~Bertone, J.~Stout and G.~M.~Tomaselli,
%``Sharp Signals of Boson Clouds in Black Hole Binary Inspirals,''
Phys. Rev. Lett. \textbf{128} (2022) no.22, 221102
%doi:10.1103/PhysRevLett.128.221102
[arXiv:2206.01212 [gr-qc]].

\bibitem{EventHorizonTelescope:2019pgp}
K.~Akiyama \textit{et al.} [Event Horizon Telescope],
%``First M87 Event Horizon Telescope Results. V. Physical Origin of the Asymmetric Ring,''
Astrophys. J. Lett. \textbf{875} (2019) no.1, L5
%doi:10.3847/2041-8213/ab0f43
[arXiv:1906.11242 [astro-ph.GA]].

\bibitem{EventHorizonTelescope:2022xnr}
K.~Akiyama \textit{et al.} [Event Horizon Telescope],
%``First Sagittarius A* Event Horizon Telescope Results. I. The Shadow of the Supermassive Black Hole in the Center of the Milky Way,''
Astrophys. J. Lett. \textbf{930} (2022) no.2, L12
%doi:10.3847/2041-8213/ac6674

\bibitem{Herdeiro:2022yle}
C.~A.~R.~Herdeiro,
%``Black holes: on the universality of the Kerr hypothesis,''
[arXiv:2204.05640 [gr-qc]].



%%%%%%%%%%%%%%%%%%%%%%%%%%%%%%%%%%%%%%%%%%%%%%%%%%%%%%

\bibitem{Hawking}
S.W. Hawking and G.F.R. Ellis, The Large Scale Structure of Spacetime (Cambridge
University Press, Cambridge 1973).


\bibitem{Sakharov:1966aja}
A.~D.~Sakharov,
%``Nachal'naia stadija rasshirenija Vselennoj i vozniknovenije neodnorodnosti raspredelenija veshchestva,''
Sov. Phys. JETP \textbf{22} (1966), 241

\bibitem{Gliner}
E. B. Gliner, Sov. Phys. JETP 22, (1966) 378


\bibitem{Bardeen}
J. M. Bardeen, “Non-singular general-relativistic gravitational collapse”,
in Proceedings of International Conference GR5, 1968, Tbilisi, USSR, p. 174.


\bibitem{Bogojevic:1998ma}
A.~Bogojevic and D.~Stojkovic,
%``A Nonsingular black hole,''
Phys. Rev. D \textbf{61} (2000), 084011
%doi:10.1103/PhysRevD.61.084011
[arXiv:gr-qc/9804070 [gr-qc]].


\bibitem{Hayward:2005gi}
S.~A.~Hayward,
%``Formation and evaporation of regular black holes,''
Phys. Rev. Lett. \textbf{96} (2006), 031103
%doi:10.1103/PhysRevLett.96.031103
[arXiv:gr-qc/0506126 [gr-qc]].

\bibitem{Bambi:2013ufa}
C.~Bambi and L.~Modesto,
%``Rotating regular black holes,''
Phys. Lett. B \textbf{721} (2013), 329-334
%doi:10.1016/j.physletb.2013.03.025
[arXiv:1302.6075 [gr-qc]].

\bibitem{Toshmatov:2014nya}
B.~Toshmatov, B.~Ahmedov, A.~Abdujabbarov and Z.~Stuchlik,
%``Rotating Regular Black Hole Solution,''
Phys. Rev. D \textbf{89} (2014) no.10, 104017
%doi:10.1103/PhysRevD.89.104017
[arXiv:1404.6443 [gr-qc]].



\bibitem{Fan:2016hvf}
Z.~Y.~Fan and X.~Wang,
%``Construction of Regular Black Holes in General Relativity,''
Phys. Rev. D \textbf{94} (2016) no.12, 124027
%doi:10.1103/PhysRevD.94.124027
[arXiv:1610.02636 [gr-qc]].



\bibitem{Simpson:2018tsi}
A.~Simpson and M.~Visser,
%``Black-bounce to traversable wormhole,''
JCAP \textbf{02} (2019), 042
%doi:10.1088/1475-7516/2019/02/042
[arXiv:1812.07114 [gr-qc]].

\bibitem{Simpson:2021dyo}
A.~Simpson and M.~Visser,
%``The eye of the storm: a regular Kerr black hole,''
JCAP \textbf{03} (2022) no.03, 011
%doi:10.1088/1475-7516/2022/03/011
[arXiv:2111.12329 [gr-qc]].

\bibitem{Simpson:2021zfl}
A.~Simpson and M.~Visser,
%``Astrophysically viable Kerr-like spacetime,''
Phys. Rev. D \textbf{105} (2022) no.6, 064065
%doi:10.1103/PhysRevD.105.064065
[arXiv:2112.04647 [gr-qc]].


\bibitem{Simpson:2019mud}
A.~Simpson and M.~Visser,
%``Regular black holes with asymptotically Minkowski cores,''
Universe \textbf{6} (2019) no.1, 8
%doi:10.3390/universe6010008
[arXiv:1911.01020 [gr-qc]].

\bibitem{Simpson:2019cer} 
A.~Simpson, P.~Martin-Moruno and M.~Visser,
%``Vaidya spacetimes, black-bounces, and traversable wormholes,''
Class. Quant. Grav. \textbf{36} (2019) no.14, 145007
%doi:10.1088/1361-6382/ab28a5
[arXiv:1902.04232 [gr-qc]].

\bibitem{Simpson:2021biv}
A.~M.~Simpson,
%``Ringing of the Regular Black Hole with Asymptotically Minkowski Core,''
Universe \textbf{7} (2021) no.11, 418
%doi:10.3390/universe7110418
[arXiv:2109.11878 [gr-qc]].

\bibitem{Franzin:2022iai}
E.~Franzin, S.~Liberati, J.~Mazza, R.~Dey and S.~Chakraborty,
%``Scalar perturbations around rotating regular black holes and wormholes: quasi-normal modes, ergoregion instability and superradiance,''
[arXiv:2201.01650 [gr-qc]].

\bibitem{Bargueno:2020ais}
P.~Bargue\~no,
%``Some global, analytical and topological properties of regular black holes,''
Phys. Rev. D \textbf{102} (2020) no.10, 104028
%doi:10.1103/PhysRevD.102.104028
[arXiv:2008.02680 [gr-qc]].


\bibitem{Berry:2020ntz}
T.~Berry, A.~Simpson and M.~Visser,
%``Photon spheres, ISCOs, and OSCOs: Astrophysical observables for regular black holes with asymptotically Minkowski cores,''
Universe \textbf{7} (2020) no.1, 2
%doi:10.3390/universe7010002
[arXiv:2008.13308 [gr-qc]].

\bibitem{Tsukamoto:2020bjm}
N.~Tsukamoto,
%``Gravitational lensing in the Simpson-Visser black-bounce spacetime in a strong deflection limit,''
Phys. Rev. D \textbf{103} (2021) no.2, 024033
%doi:10.1103/PhysRevD.103.024033
[arXiv:2011.03932 [gr-qc]].

\bibitem{Zhou:2020zys}
T.~Y.~Zhou and Y.~Xie,
%``Precessing and periodic motions around a black-bounce/traversable wormhole,''
Eur. Phys. J. C \textbf{80} (2020) no.11, 1070
%doi:10.1140/epjc/s10052-020-08661-w

\bibitem{Shaikh:2021yux}
R.~Shaikh, K.~Pal, K.~Pal and T.~Sarkar,
%``Constraining alternatives to the Kerr black hole,''
Mon. Not. Roy. Astron. Soc. \textbf{506} (2021) no.1, 1229-1236
%doi:10.1093/mnras/stab1779
[arXiv:2102.04299 [gr-qc]].

\bibitem{Mazza:2021rgq}
J.~Mazza, E.~Franzin and S.~Liberati,
%``A novel family of rotating black hole mimickers,''
JCAP \textbf{04} (2021), 082
%doi:10.1088/1475-7516/2021/04/082
[arXiv:2102.01105 [gr-qc]].

\bibitem{Lima:2021las}
H.~C.~D.~Lima, Junior., L.~C.~B.~Crispino, P.~V.~P.~Cunha and C.~A.~R.~Herdeiro,
%``Can different black holes cast the same shadow?,''
Phys. Rev. D \textbf{103} (2021) no.8, 084040
%doi:10.1103/PhysRevD.103.084040
[arXiv:2102.07034 [gr-qc]].

\bibitem{Islam:2021ful}
S.~U.~Islam, J.~Kumar and S.~G.~Ghosh,
%``Strong gravitational lensing by
rotating Simpson-Visser black holes,''
JCAP \textbf{10} (2021), 013
%doi:10.1088/1475-7516/2021/10/013
[arXiv:2104.00696 [gr-qc]].

\bibitem{Bueno:2021krl}
P.~Bueno, P.~A.~Cano, J.~Moreno and G.~van der Velde,
%``Regular black holes in three dimensions,''
Phys. Rev. D \textbf{104} (2021) no.2, L021501
%doi:10.1103/PhysRevD.104.L021501
[arXiv:2104.10172 [gr-qc]].

\bibitem{Giacchini:2021pmr}
B.~L.~Giacchini, T.~d.~Netto and L.~Modesto,
%``Action principle selection of regular black holes,''
Phys. Rev. D \textbf{104} (2021) no.8, 084072
%doi:10.1103/PhysRevD.104.084072
[arXiv:2105.00300 [gr-qc]].

\bibitem{Stuchlik:2021tcn}
Z.~Stuchl\'\i{}k and J.~Vrba,
%``Epicyclic Oscillations around Simpson\textendash{}Visser Regular Black Holes and Wormholes,''
Universe \textbf{7} (2021) no.8, 279
%doi:10.3390/universe7080279
[arXiv:2108.09562 [gr-qc]].

\bibitem{Cheng:2021hoc}
X.~T.~Cheng and Y.~Xie,
%``Probing a black-bounce, traversable wormhole with weak deflection gravitational lensing,''
Phys. Rev. D \textbf{103} (2021) no.6, 064040
%doi:10.1103/PhysRevD.103.064040

\bibitem{Patel:2022jbk}
V.~Patel, K.~Acharya, P.~Bambhaniya and P.~S.~Joshi,
%``Rotational energy extraction from the Kerr black hole's mimickers,''
[arXiv:2206.00428 [gr-qc]].

\bibitem{Ghosh:2014pba}
  S.~G.~Ghosh,
  %``A nonsingular rotating black hole,''
  Eur. Phys. J. C \textbf{75} (2015) no.11, 532
 %doi:10.1140/epjc/s10052-015-3740-y
  [arXiv:1408.5668 [gr-qc]].
  
  
\bibitem{Detweiler:1980uk}
 S.~L.~Detweiler,
 %``KLEIN-GORDON EQUATION AND ROTATING BLACK HOLES,''
 Phys. Rev. D \textbf{22} (1980), 2323-2326
 %doi:10.1103/PhysRevD.22.2323 
 
\bibitem{Herdeiro:2014jaa}
C.~Herdeiro and E.~Radu,
%``Ergosurfaces for Kerr black holes with scalar hair,''
Phys. Rev. D \textbf{89} (2014) no.12, 124018
%doi:10.1103/PhysRevD.89.124018
[arXiv:1406.1225 [gr-qc]]. 

\end{thebibliography}
\end{document}